\documentclass[draft,jgrga]{agutex}
\usepackage{lineno}

\usepackage{graphicx}
\usepackage{epstopdf}
\usepackage{color}
\usepackage{url}

\def\bw{{\cal B}_w}
\def\s{{\cal S}}

\newcommand{\ssr}{    {Space Sci. Rev.}}
\newcommand{\planss}{    {Plan. Space Sci.}}
\newcommand{\nat}{    {Nature}}
\newcommand{\solphys}{    {Solar Phys.}}

\setkeys{Gin}{draft=false}

\authorrunninghead{VAINCHTEIN ET AL.}
\titlerunninghead{Electron Nonlinear Resonances with intense chorus waves}
\authoraddr{Vainchtein D., Nyheim Plasma Institute, Drexel University, Camden, NJ, USA (dlv36@drexel.edu)}
\authoraddr{Zhang X.-J., Department of Earth, Planetary, and Space Sciences, University of California, Los Angeles, USA (xjzhang@ucla.edu)}
\authoraddr{Artemyev A. V., Department of Earth, Planetary, and Space Sciences, University of California, Los Angeles, USA (aartemyev@igpp.ucla.edu)}
\authoraddr{Mourenas D., CEA, DAM, DIF, F 91297 Arpajon Cedex, France}
\authoraddr{Angelopoulos V., Department of Earth, Planetary, and Space Sciences, University of California, Los Angeles, USA}
\authoraddr{Thorne R. M., Department of Atmospheric and Oceanic Sciences, University of California, Los Angeles, USA}

\begin{document}

\title{Evolution of electron distribution driven by nonlinear resonances with intense field-aligned chorus waves}



\authors{D. Vainchtein\altaffilmark{1,4}, X.-J. Zhang\altaffilmark{2,3}, A. V. Artemyev\altaffilmark{2,4}, D. Mourenas\altaffilmark{5}, \\
V. Angelopoulos\altaffilmark{2}, R. M. Thorne\altaffilmark{3}}

\altaffiltext{1}{Nyheim Plasma Institute, Drexel University, Camden, NJ, USA}
\altaffiltext{2}{Department of Earth, Planetary, and Space Sciences, University of California, Los Angeles, CA, USA}
\altaffiltext{3}{Department of Atmospheric and Oceanic Sciences, University of California, Los Angeles, CA, USA}
\altaffiltext{4}{Space Research Institute, Russian Academy of Sciences, Moscow, Russia}
\altaffiltext{5}{CEA, DAM, DIF, Arpajon, France}

\begin{abstract}
Resonant electron interaction with whistler-mode chorus waves is recognized as one of the main drivers of radiation belt dynamics. For moderate wave intensity, this interaction is well described by quasi-linear theory. However, recent statistics of parallel propagating chorus waves have demonstrated that $5-20$\% of the observed waves are sufficiently intense to interact nonlinearly with electrons. Such interactions include phase
trapping and phase bunching (nonlinear scattering) effects not described by the quasi-linear diffusion. For sufficiently long (large) wave-packets, these nonlinear effects can result in very rapid electron acceleration and scattering. In this paper we introduce a method to include trapping and nonlinear scattering into the kinetic equation describing the evolution of the electron distribution function. We use statistics of Van Allen Probes and Time History of Events and Macroscale Interactions during Substorms (THEMIS) observations to determine the probability distribution of intense, long wave-packets as function of power and frequency. Then we develop an analytical model of particle resonance of an individual particle with an intense chorus wave-packet and derive the main properties of this interaction: probability of electron trapping, energy change due to trapping and nonlinear scattering. These properties are combined in a nonlocal operator acting on the electron distribution function. When multiple waves are present, we average the obtained operator over the observed distributions of waves and examine solutions of the resultant kinetic equation. We also examine energy conservation and its implications  in systems with the nonlinear wave-particle interaction.
\end{abstract}

\begin{article}

{\bf Key Points}:
\begin{enumerate}
\item {\it We propose a model of electron nonlinear resonant interaction with long and intense chorus wave-packets}
\item {\it We derive a new generalized kinetic equation for electrons that encompasses nonlinear interactions with long chorus wave-packets}
\item {\it Nonlinear interactions with long wave-packets can produce rapid electron acceleration for observed wave characteristics}\\
\end{enumerate}

\section{Introduction}
The dynamics of Earth's radiation belts is partly controlled by efficient resonant interactions of electrons with whistler-mode waves \citep{Andronov&Trakhtengerts64,Kennel&Petschek66}. Different modes of these waves
are responsible for warm electron precipitation and aurora formation \citep[lower and upper bands chorus waves, see, e.g.,][]{Thorne10:Nature,Ni16:ssr,Kasahara18:nature}, electron acceleration and formation of
relativistic electron fluxes in the heart of the outer radiation belt \citep[lower band chorus waves, see, e.g.,][]{Thorne13:nature,Reeves13,Li14:storm}, electron scattering and flux reduction during radial diffusion
\citep[hiss waves, see, e.g.,][]{Ma16:hiss, Mourenas17}, electron scattering in the inner radiation belt \citep[VLF waves from ground-based transmitters, see, e.g.,][]{Subbotin11, Agapitov14:jgr:AKEBONO, Ma17:vlf}.
Depending on their actual characteristics, such whistler-mode waves can resonantly interact with electrons over a wide range of energies starting from $100$ eV \citep[very oblique chorus waves, see,
e.g.,][]{Agapitov15:jgr, Artemyev15:natcom} and up to several MeVs \citep[intense parallel propagating chorus waves, see, e.g.,][]{Omura07}. The wave frequency and obliquity define electron resonant energies
\citep[see][and references therein]{Shklyar09:review, Artemyev16:ssr} and the wave intensity determines the regime of resonant interaction. Low intensity waves provide electron diffusive scattering, traditionally
described by quasi-linear theory \citep[][]{Vedenov62,Kennel&Engelmann66}. In the inhomogeneous geomagnetic field of the radiation belts, the applicability of quasi-linear theory can be justified even for narrow band
waves \citep[e.g.,][]{Karpman74:ssr, Karpman&Shkliar77, Albert10} and it has been used to successfully describe many observed phenomena \citep[e.g.,][]{Thorne13:nature,  Su14, Glauert14, Ma15, Drozdov15, Li16:jgr,
Albert16}.

Unlike low intensity waves whose effects on particles have been modeled successfully by diffusion, the consequences of interaction of particles with  intense whistler-mode waves for the dynamics of the radiation belts have not yet been fully investigated. There is a common understanding that the generation
of intense whistler-mode waves is controlled by nonlinear wave-particle interaction \citep[see reviews][and references therein]{Shklyar09:review, Omura13:AGU}, including electron trapping into resonance and a significant
modification of the resonant electron distribution function \citep[][]{Nunn74, Karpman74}. The theory of nonlinear wave generation reproduces many observed properties of intense whistler-mode waves really well
\citep[e.g.,][]{Omura08, Katoh&Omura07, Demekhov11, Demekhov17, Tao17:generation, Shklyar17}. However, the exact role played by the most intense whistler-mode waves in electron acceleration still remains somewhat
controversial. Electron trapping \citep[][]{Nunn71} can provide a very long resonant interaction (over a fraction of the electron bounce period along magnetic field line), corresponding to a significant change of particle
energy \citep[ comparable to the initial particle energy, see][]{Omura07, Summers&Omura07, Bortnik08, Artemyev12:pop:nondiffusion, Tao13, Agapitov14:jgr:acceleration}. However, the effects of trapping
should be almost compensated by resonant nonlinear scattering, which provides electron drift in energy space in the opposite direction to trapping acceleration \citep[see][and references therein]{Shklyar11:angeo}.
Although the fine balance between trapping and nonlinear scattering is expected to ultimately establish a kind of diffusive regime of particle acceleration \citep[see][]{Shklyar81, Solovev&Shkliar86}, the actual
time required for establishing such a regime can be very long. Thus, when trapping and nonlinear scattering processes do not exactly balance each other, they can potentially provide a very rapid (compared with
quasi-linear diffusion) net acceleration of electrons \citep[e.g.,][]{Demekhov09, Nunn&Omura15, Hsieh&Omura17}.  Models of such rapid acceleration have been able to explain some observations of fast increase of energetic
electron flux \citep[e.g.,][]{Agapitov15:grl:acceleration, Mozer16, Foster17}, but the global importance of such fast acceleration processes in the overall radiation belt dynamics remains unknown.

To include the nonlinear resonant interactions into global models of wave-particle interaction in the radiation belts one needs: (i) representative statistics of intense wave characteristics based on spacecraft measurements and (ii) a universal approach for the description of the evolution of the electron distribution due to nonlinear interactions. There is a lack of information about fine wave characteristics such as the intensity and duration of individual wave-packets. Because most wave statistics are collected for use in quasi-linear diffusion codes, their
focus is on the average wave intensity, over long time intervals \citep[see, e.g.,][]{Meredith12, Li13, Agapitov13:jgr}. Such averaging mixes two different wave
populations, low intensity waves and transient bursts of high intensity waves, and it does not allow to consider them separately afterwards. Thus, most wave statistics do not contain any information about wave-packet
characteristics, such as the duration of wave-packets (or wave-packet modulation), which are critical for modeling nonlinear wave-particle interaction \citep[][]{Tao12, Tao13, Artemyev12:pop:nondiffusion} but not important for quasi-linear models. Recently, \citet{Zhang18:jgr:intensewaves} analyzed THEMIS and Van Allen Probe measurements of intense parallel chorus whistler-mode waves and identified the portion of the
wave population that may potentially interact with electrons in the nonlinear regime. Most of intense chorus waves ($\sim 95-99$\%) were found to propagate in the form of short wave-packets, allowing resonant interaction
with electrons over only a very short time, despite their large intensity \citep{Zhang18:jgr:intensewaves, Mourenas18:jgr}. Only a few the intense waves propagate in the form of long wave-packets and can efficiently accelerate trapped electrons. In this study we use the statistics collected by \citet{Zhang18:jgr:intensewaves} to quantify how these few percents of intense wave-packets that are sufficiently long to produce a significant nonlinear acceleration affect the electron distribution.

\begin{figure*}
\includegraphics[width=0.8\textwidth]{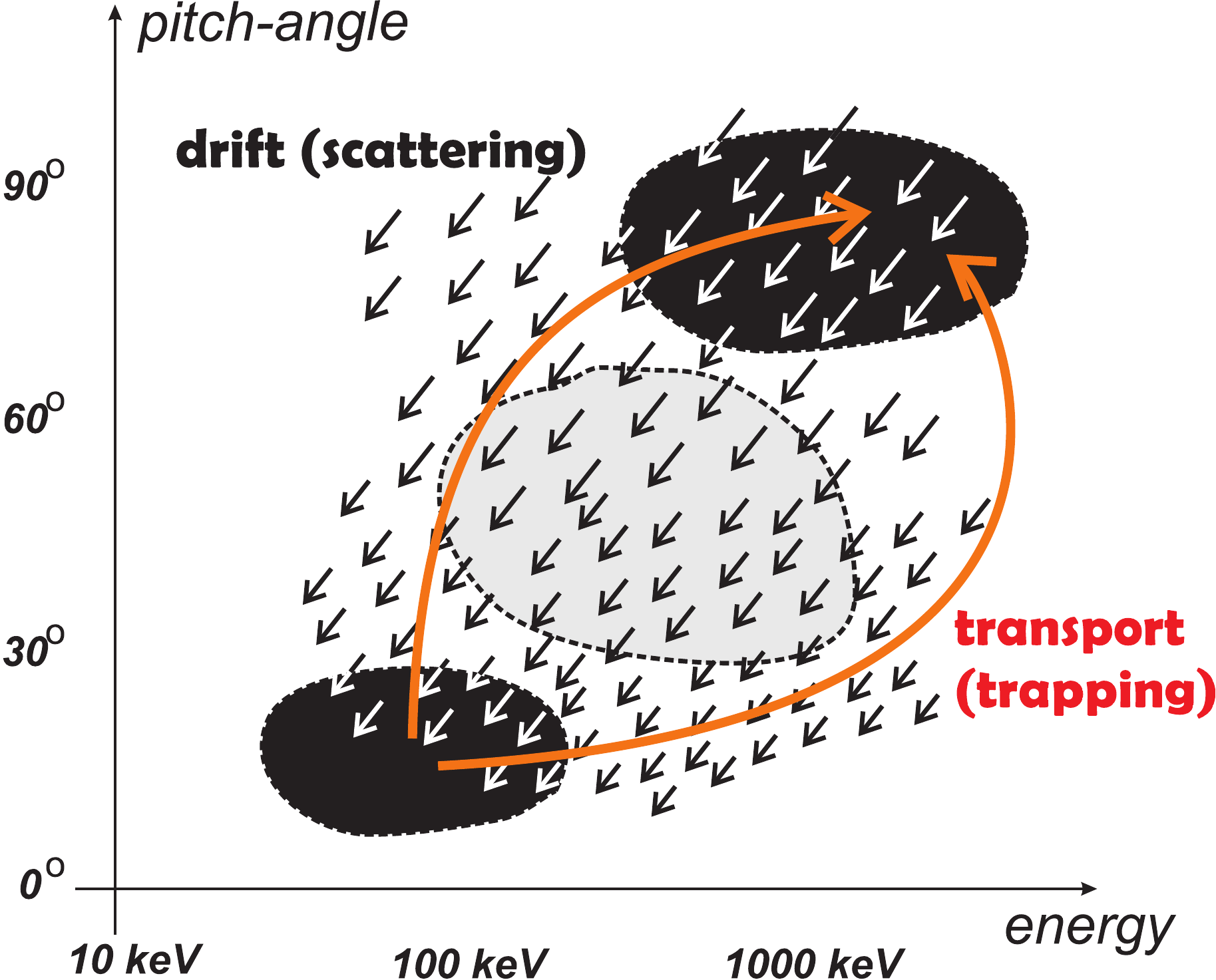}
\centering
\caption{A schematic view of particle transport in phase space. The grey color indicates the parametric area of the expected phase space density depletion: particles are scattered from this region to smaller energy/pitch-angle whereas particle scattered into this region come from a lower phase space density region. Black color indicates the areas of the expected phase space density increase: particles are scattered (to lower energies) or transported by trapping (to higher energies) into these regions. Two red arrows connecting these different areas show that trapping acceleration is not a local process.}
\label{fig1}
\end{figure*}

Although the test particle approach provides a satisfactory description of the effects of the nonlinear resonant interactions on the electron distribution \citep[][]{Bortnik08, Yoon13:whistlers, Nunn&Omura15, Agapitov16:grl},
this approach cannot be used to calculate the long-term dynamics of the radiation belts. Therefore, theoretical estimates of electron acceleration/deceleration due to nonlinear trapping and scattering
\citep[e.g.,][]{Shklyar81, Bell84, Albert93} should be combined in some operator acting on the full electron distribution function to complement and generalize the usual quasi-linear diffusion equation. Several examples
of such operators were developed numerically \citep[e.g.,][]{Omura15, Hsieh&Omura17:radio_science} and analytically in the two opposite limiting cases of infinitely long wave-packets \citep[e.g.,][]{Artemyev14:grl:fast_transport, Artemyev16:pop:letter} and short wave-packets \citep{Mourenas18:jgr}. However, for long but finite wave-packets, one should consider realistic distributions of intense wave characteristics. Nonlinear operators describing fast particle transport in energy space due to trapping contain nonlocal terms describing transport between different distant
energy/pitch-angle domains (see schematic  in Fig. \ref{fig1}). Averaging such nonlocal operators over the distribution of observed wave characteristics is therefore a difficult, still unsolved problem.

In this study, we attempt to quantify the effects of the nonlinear wave-particle interaction on the evolution of the global electron distribution. First we use statistics of intense, long wave-packet chorus waves collected by
\citet{Zhang18:jgr:intensewaves} to determine their probability distribution as function of wave intensity and frequency (Sect. \ref{sec:statistics}). Then we consider the nonlinear wave-particle interaction and determine its main characteristics (Sect. \ref{sec:resonances}). These characteristics are used to construct a generalized kinetic equation including the effects of nonlinear wave-particle interactions with long
wave-packets (Sect. \ref{sec:kinetics}). The operators of this equation are averaged over the observed distribution of wave characteristics. Then, we demonstrate how the averaged nonlinear effects drive the evolution of
the electron distribution (Sect. \ref{sec:distribution}). Finally, we discuss the applicability of the proposed approach for the description of the actual dynamics of the radiation belts and the particular constraints on
the electron distribution that follow from the obtained solutions of the generalized kinetic equation (Sect. \ref{sec:discussion}).

\section{Intense whistler-mode wave characteristics \label{sec:statistics}}
Two wave characteristics that are the most important for the nonlinear wave-particle interaction are the wave amplitude $B_w$ (square root of wave intensity) and the duration of the wave-packet, given here by the number
$\beta$ of wave periods within one packet. The normalized wave amplitude is $\bw=R\Omega_{ce}B_w/cB_0$, where $B_0$ is the equatorial background magnetic field, $\Omega_{ce}$ the equatorial electron gyrofrequency,
$R=R_EL$ the spatial scale of the background magnetic field inhomogeneity, and $c$ the speed of light. If $\bw >2$, nonlinear interaction is possible, while for $\bw<1-2$ the wave is not sufficiently intense to trap particles  \citep[for more details, see][and references therein]{Zhang18:jgr:intensewaves,Mourenas18:jgr}. A more precise evaluation of the critical wave amplitude necessary for the nonlinear interaction depends on
additional wave characteristics (e.g., frequency drift) and resonant particle energy and pitch-angle \citep[see, e.g.,][]{Karpman74:ssr, Bell86, LeQueau&Roux87}, but a aforementioned simplified $\bw$-criteria is
sufficiently accurate to process the observed wave statistics and provide first-order estimates of the impact on the electron distribution. We use lower-band chorus wave statistics collected by
\citet{Zhang18:jgr:intensewaves} for $\bw>2$. Five years of THEMIS \citep{Angelopoulos08:ssr} and three years of Van Allen Probes \citep{Mauk13} wave measurements have been analyzed to identify intense chorus
wave-packets and determine their characteristics. Onboard THEMIS, wave fields are measured by search coil magnetometers \citep{LeContel08} and the electric field instrument \citep{Bonnell08}. We also use measurements of
three components of the background magnetic field by the fluxgate magnetometer \citep{Auster08:THEMIS}. The two identical Van Allen Probes measure electric and magnetic field waveforms using the Electric Fields and Waves
\citep{Wygant13} and the Electric and Magnetic Field Instrument Suite and Integrated Science \citep{Kletzing13} detectors. The data is transmitted at $35000$ samples/s over $6$s intervals in burst mode.

\begin{figure*}
\includegraphics[width=0.9\textwidth]{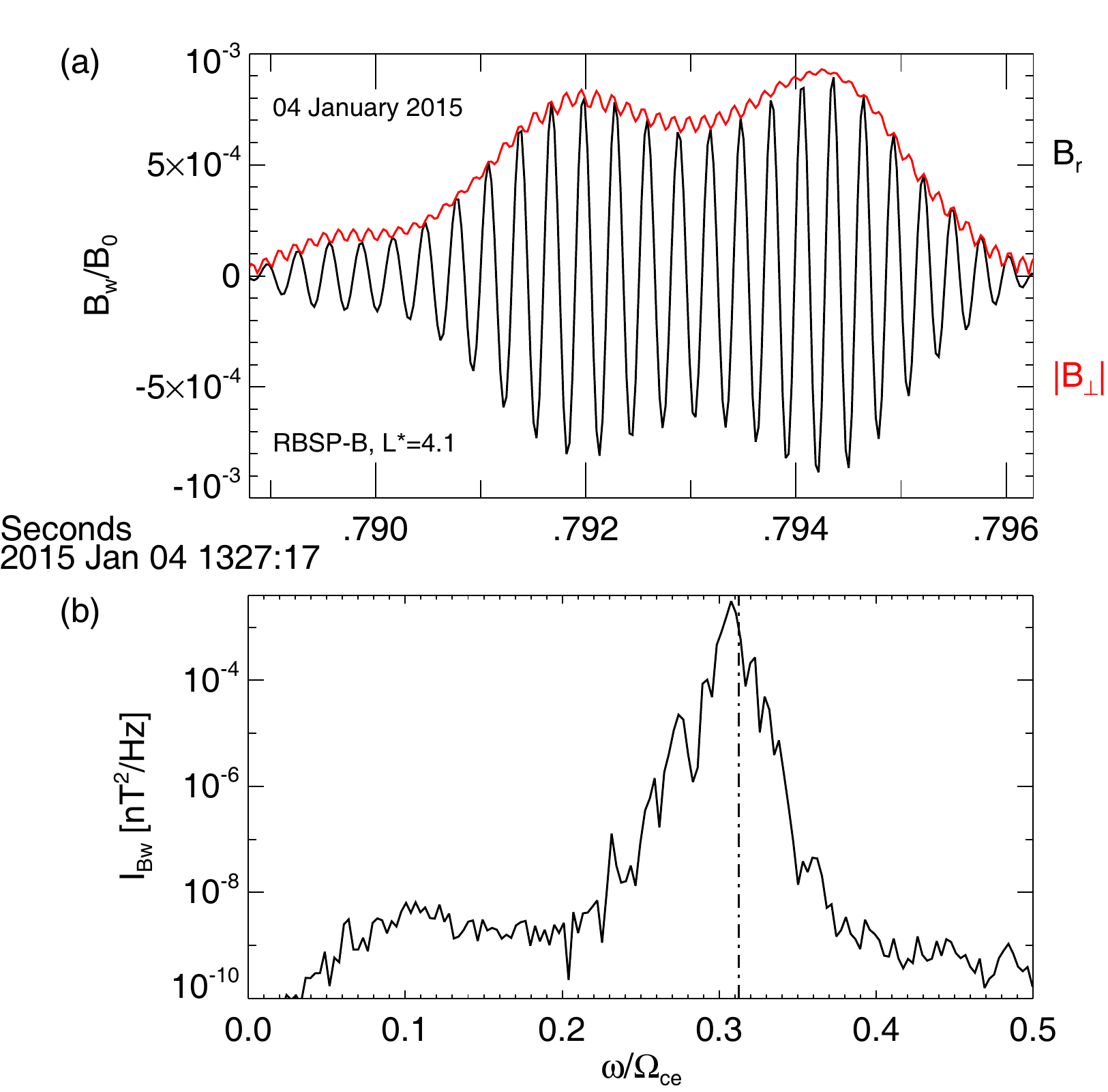}
\centering
\caption{Example of long wave-packets, wave-packet, $B_\bot$ and $B_\bot/B_0$, spectrum with $\omega/\Omega_{ce}$.}
\label{fig2}
\end{figure*}

Due to the difference between THEMIS and Van Allen Probe operational modes (the absolute majority of the THEMIS waveforms measurements are triggered by plasma injections), the occurrence rate of $\bw>2$ waves are
different for datasets collected by THEMIS and the Van Allen Probes, but on average there are $<1$\% of intense ($\bw>2$) parallel chorus waves propagating along magnetic field lines in the form of long wave-packets (with
$\beta\geq 50$). One example of a long wave-packet is shown in Figure~\ref{fig2}. The main wave parameters, $B_w$ and frequency $\omega$, are defined for each such packet and normalized on the background
characteristics: $B_w/B_0$ and $\omega/\Omega_{ce}$ where $B_0$ and $\Omega_{ce}$ are evaluated using the equatorial geomagnetic field. Because most of THEMIS and Van Allen Probe measurements are not equatorial, we use
$L^*$ for Van Allen Probe data to approximate the equatorial field and $B_z$ GSM measurements of THEMIS (at large $L$-shell) as a proxy of the equatorial field. All THEMIS measurements are restricted to the
near-equatorial region with $B_z>\sqrt{B_x^2+B_y^2}$.

\begin{figure*}
\includegraphics[width=0.9\textwidth]{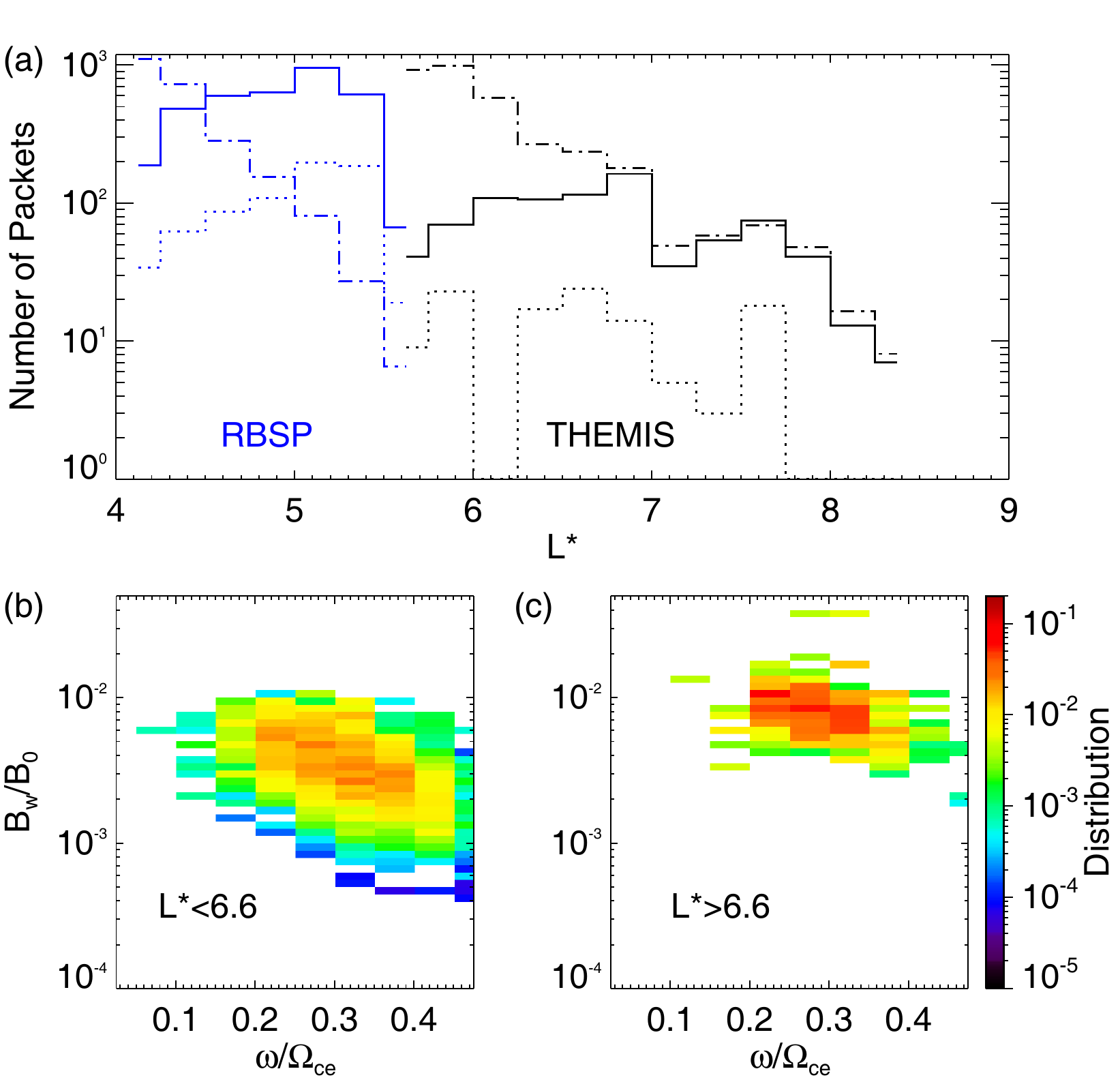}
\centering
\caption{(a) Per day observations of wave-packets of intense chorus wave as a function of $L^*$ for $\beta>50$ (solid lines) and $\beta^*>50$ (dotted lines), (b), (c) distributions in $B_w,
\omega/\Omega_{ce}(0)$ for $\bw>2$ and $\beta>50$, for two $L^*$-shell ranges.}
\label{fig3}
\end{figure*}

The number of chorus wave-packets with $\bw>2$ and $\beta>50$ captured by THEMIS and the Van Allen Probes per day of intense chorus wave measurements are shown in Fig.~\ref{fig3}(a) as a function of $L$-shell. Van Allen Probes ($L<6$) captured much more wave-packets than THEMIS ($L>6$), but the occurrence rate of intense waves is larger at larger $L$ (where THEMIS spacecraft provide most of the statistics). This is mainly due to the shorter total time of waveform measurements by THEMIS \citep[see details in][]{Zhang18:jgr:intensewaves}. The parameter $\beta^*$ is number of wave-periods with wave magnetic field larger than half of the peak wave-packet magnetic field. The distribution of the number of observed wave-packets with $\beta^*>50$ (dotted lines in Fig.~\ref{fig3}(a)) shows that packets with $\beta^*>50$ represent only $\sim 15-20$\% of all long
wave-packets with $\beta>50$. The remaining $\sim 80-85$\% of these long wave-packets are actually still very localized, i.e., most of the wave intensity is located near the position of peak intensity, while the
remaining part (the tails) of the wave-packet is much less intense. This factor can be important for realistic estimates of the efficiency of nonlinear wave-particle interaction in the radiation belts.

We separate the $L^*$-shell range into two intervals: (1) the outer radiation belt with $L^*$ between the plasmasphere \citep[as defined by the model from][]{OBrien&Moldwin03} and the geostationary orbit $L^*\sim 6.6$, and (2) the injection region at $L^*\in [6.6, 9]$. These two $L^*$-shell ranges roughly separate Van Allen Probe measurements (outer radiation belt) and THEMIS measurements (injection region). For both $L^*$-shell ranges, we
plot the wave-packet distributions in the $(B_w/B_0, \omega/\Omega_{ce})$ space. Figures \ref{fig3}(b) and (c) show the presence of a significant portion of intense ($B_w/B_0\in [10^{-3},10^{-2}]$) waves with $\omega/\Omega_{ce}\in[0.2,0.4]$. We use these wave distributions to estimate
the average effect of the nonlinear wave-particle interaction on the evolution of the electron distribution.

\section{Nonlinear resonances \label{sec:resonances}}
\begin{figure*}
\includegraphics[width=0.9\textwidth]{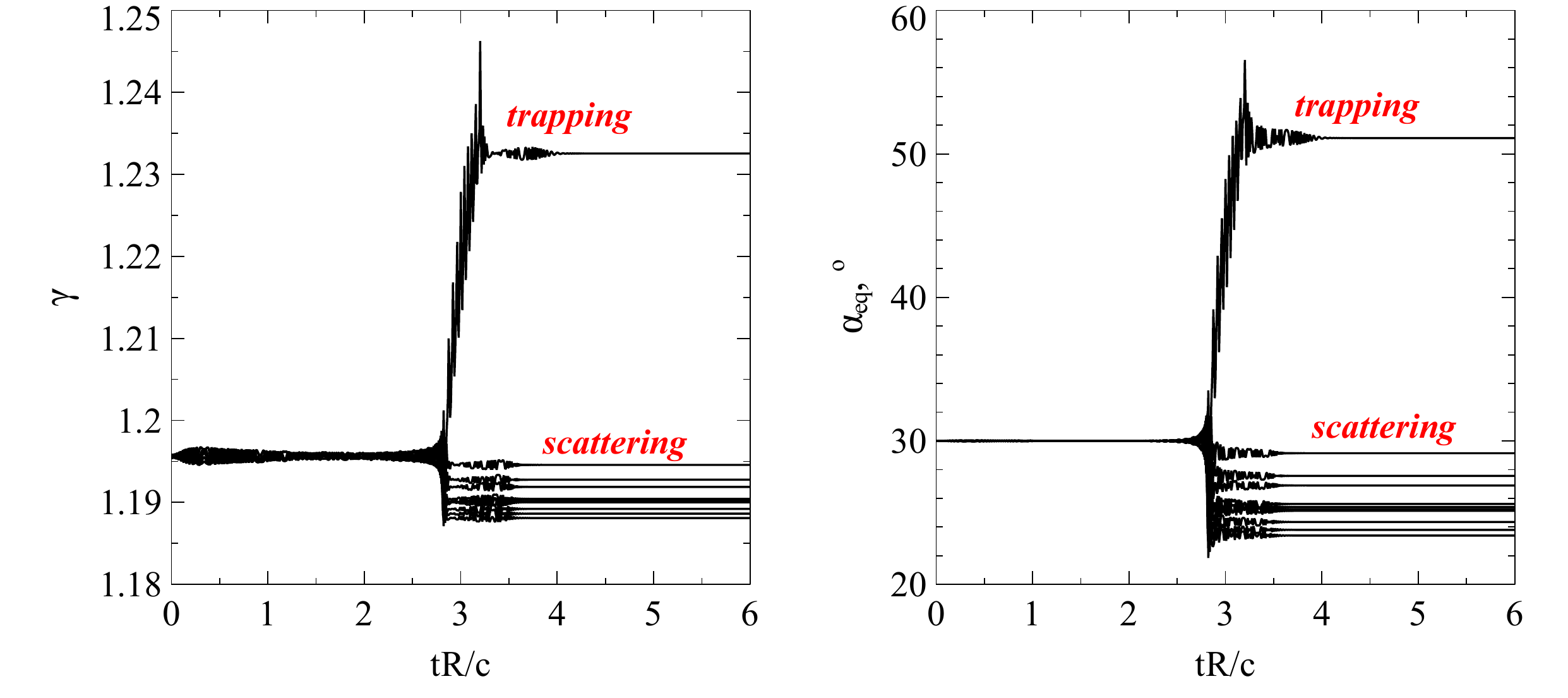}
\centering
\caption{Results of numerical integration of 10 particle trajectories described by Eqs.~(\ref{eq02a}). All particles have the same initial energy and equatorial pitch-angle (initial $I_x$), and different (randomly
distributed) initial phases $\psi$. Panels show energy (left panel) and equatorial pitch-angle (right panel) as a function of dimensionless time (trajectories are integrated for $\sim$ a quarter of the bounce period;
there is only one resonance for each trajectory). System parameters are: $L=6$, $\omega/\Omega_{ce}(0)=0.35$, $B_w=500{\rm pT}\cdot \tanh(\lambda/5^{\circ})\exp{-(\lambda/25^{\circ})^2}$ for $\lambda>0$ \citep[i.e., the
latitudinal distribution used  the mimics obsereved wave field distribution, see][]{Agapitov13:jgr}.}
\label{fig4}
\end{figure*}

We consider a simple planar magnetic field model \citep{Bell84} with a single vector potential component $A_{0y}=-xB_0(z)$, where $z$ is a field-aligned coordinate and $B_0(z)$ mimics the dipolar
geomagnetic field ($B_0=\sqrt{1+3\sin^2\lambda}/\cos^6\lambda$ where $dz/d\lambda = R\sqrt{1+3\sin^2\lambda}$ and $R=R_EL$). The parallel propagating whistler-mode wave is described by two components of the vector
potential: $A_x=(B_w/k)\sin\phi$ and $A_y=(B_w/k)\cos\phi$ where $B_w$ and $k$ are the wave amplitude and wave vector ($B_w$ and $k$ depend on $z/R$), and $\phi$ is the wave phase ($\partial \phi/\partial z=k$, $\partial
\phi/\partial t=-\omega$). The Hamiltonian of a relativistic electron (the charge being $-e$ and the rest mass $m_e$) in such electromagnetic fields is
\begin{linenomath}
\begin{equation}
H = \sqrt {m_e^2 c^4  + c^2 p_z^2  + \left( {cp_x  + eA_x } \right)^2  + e^2 \left( {A_{0y}  + A_y } \right)^2 }
\label{eq01}
\end{equation}
\end{linenomath}
We expand Hamiltonian (\ref{eq01}) over a small parameter $eB_w/km_ec^2\ll 1$ and introduce new conjugate variables, the gyrophase $\psi$ and magnetic moment $I_x=\oint{p_xdx}$ \citep[see details of Hamiltonian
transformation in, e.g.,][]{Artemyev15:pop:probability, Artemyev18:jpp}. The new Hamiltonian takes the form:
\begin{linenomath}
\begin{equation}
H = m_e c^2 \gamma  + U_w (z,I_z )\sin \left( {\phi  + \psi } \right),\quad \gamma  = \sqrt {1 + \frac{{p_z^2 }}{{m_e^2 c^2 }} + \frac{{2I_x \Omega _{ce} }}{{m_e c^2 }}}
\label{eq02}
\end{equation}
\end{linenomath}
where $\gamma$ is the gamma factor of the gyro-averaged system, $\Omega_{ce}=eB_0(z)/m_ec$, and $U_w=\sqrt{2I_x\Omega_{ce}}eB_w/\gamma m_eck$ is the effective wave amplitude. The Hamiltonian equations
describe electron motion:
\begin{linenomath}
\begin{eqnarray}
 \dot z &=& \frac{{\partial H}}{{\partial p_z }} = \frac{{p_z }}{{\gamma m_e }}+ \frac{{\partial U_w }}{{\partial p_z }}\sin \left( {\phi  + \psi } \right) \nonumber \\
 \dot \psi  &=& \frac{{\partial H}}{{\partial I_x }} = \frac{{\Omega _{ce} }}{\gamma } + \frac{{\partial U_w }}{{\partial I_x }}\sin \left( {\phi  + \psi } \right) \nonumber \\
 \dot p_z  &=&  - \frac{{\partial H}}{{\partial z}} =  - \frac{{I_x \Omega '_{ce} }}{\gamma } - kU_w \cos \left( {\phi  + \psi } \right) + \frac{{\partial U_w }}{{\partial z}}\sin \left( {\phi  + \psi } \right) \nonumber
 \\
 \dot I_x  &=&  - \frac{{\partial H}}{{\partial \psi }} =  - U_w \cos \left( {\phi  + \psi } \right),\quad \dot \phi  = k\dot z - \omega  \label{eq02a}
 \end{eqnarray}
\end{linenomath}
where $\Omega_{ce}'=d\Omega_{ce}/dz$. In the absence of waves ($U_w=0$), electrons move along the bounce trajectory $\dot z=p_z/\gamma m_e$, $\dot p_z = -I_x\Omega_{ce}'/\gamma$ with a constant energy $\gamma$ and magnetic
moment $I_x$. Waves disturb these bounce oscillations and can scatter or trap particles (see Fig.~\ref{fig4}). For sufficiently intense waves, as considered in this study, nonlinear scattering results in energy change
with a finite mean value $\langle\Delta\gamma \rangle = \Delta \gamma_{scat}\ne 0$ \citep[see details in, e.g.,][]{Solovev&Shkliar86,Albert02}. Energy changes due to trapping and nonlinear scattering significantly exceed
the square root of energy variance (see relations between $\Delta \gamma_{scat}$, $\Delta \gamma_{trap}$ and energy spread in Fig.~\ref{fig4}), i.e., trapping and nonlinear scattering are the dominant processes, whereas
energy diffusion is much weaker. The diffusion process likely becomes important on very long time intervals, but it can be omitted over relatively short intervals (still including many resonant interactions). Therefore,
we focus here on the description of trapping and nonlinear scattering of electrons and on the effects of these processes on the evolution of the full electron distribution function.

Three main characteristics define the resonant interaction: (i) the probability of trapping, $\Pi$, gives the relative number of resonant particles trapped during one resonant interaction (the relative number of scattered particles is equal to $1-\Pi$); (ii) the energy change due to trapping $\Delta \gamma_{trap}$; (iii) the energy change due to nonlinear scattering $\Delta \gamma_{scat}$. These characteristics depend on the initial particle energy $\gamma$ and on the initial value of $I_x$. $I_x$ can be written through the energy and equatorial pitch-angle $\alpha_{eq}$ as
$I_x=m_ec^2(\gamma^2-1)\sin^2\alpha_{eq}/2\Omega_{ce}(0)$. Thus, we use the notation $\Pi(\gamma,\alpha_{eq})$, $\Delta\gamma_{scat}(\gamma,\alpha_{eq})$, $\Delta\gamma_{trap}(\gamma,\alpha_{eq})$. These characteristics can be combined to construct a nonlocal (integral) operator acting on the full electron distribution function to describe its evolution due to the nonlinear wave-particle interaction \citep[e.g.,][]{Solovev&Shkliar86, Artemyev16:pop:letter, Omura15}.

\begin{figure*}
\includegraphics[width=0.9\textwidth]{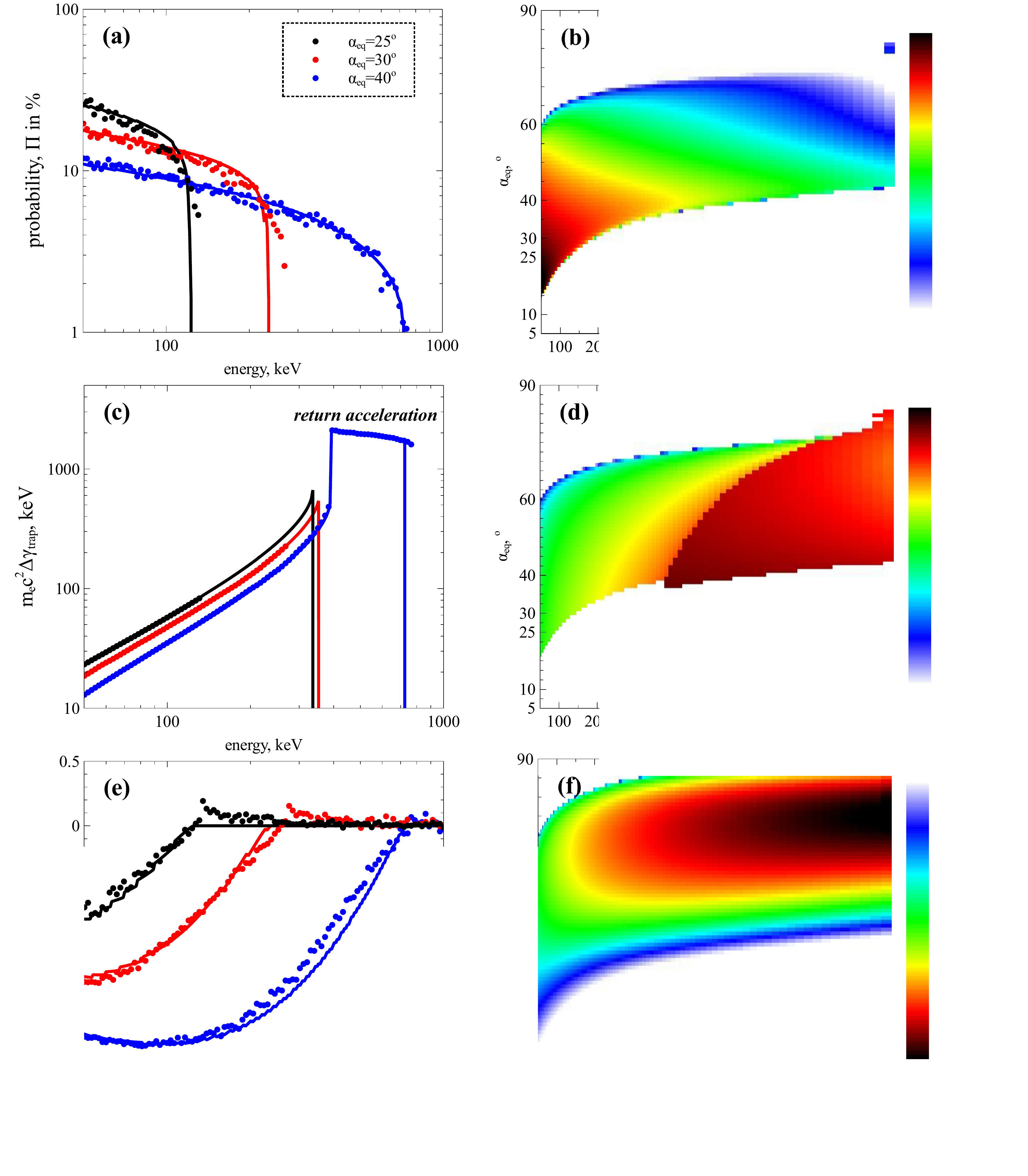}
\centering
\caption{Panels (a), (c), and (e): Probability of trapping, energy change of trapped particles, and energy change of nonlinearly scattered particles as functions of particle initial energy for three initial
pitch-angles. Curves are analytical results and circles are results of numerical integration of Hamiltonian equations (\ref{eq02a}); each circle representing values averaged over $5000$ trajectories. Panels (b), (d), and (f): Analytically evaluated probability of trapping, energy change of trapped particles, and energy change of scattered particles. System parameters are the same as in Fig.~\ref{fig4}. In panels (c) and
(d) we indicate the parameter region of the turning acceleration characterized by a more significant energy gain \citep[see details in][]{Omura07}.}
\label{fig5}
\end{figure*}

The probability of trapping, $\Pi$, can be obtained numerically using Hamiltonian equations (\ref{eq02a}). We bin $(\gamma,\alpha_{eq})$ space, and for each $(\gamma,\alpha_{eq})$ we calculated trajectories of $5\cdot10^3$ particles with the same initial position $z=0$ and different randomly distributed phases $\psi$ (initial $p_z$ and $I_x$ are defined by $(\gamma,\alpha_{eq})$). Each trajectory is integrated for a half of
the bounce period. The final $(\gamma,\alpha_{eq})$ determines whether this particle was trapped (increase of $\Delta \gamma$ exceeding $10$\% of the initial $\gamma$) or scattered ($\Delta\gamma<0$). The relative number of trapped particles is shown in Fig.~\ref{fig5}(a) for one particular set of system parameters (see the figure caption). Although such a calculation of $\Pi$ is relatively straightforward, it requires a lot of test particles with sufficiently small bins in $(\gamma,\alpha_{eq})$ space. Such a procedure should be repeated for all sets of the system parameters (wave amplitudes, wave frequencies, etc.). Thus a purely numerical determination of $\Pi$ is not very effective. Alternatively, $\Pi$ can be obtained analytically \citep[e.g.,][]{Neishtadt75, Shklyar81, Neishtadt89}. We use here the approach developed
in \citet{Artemyev15:pop:probability} (see details of $\Pi$ calculations in Appendix A, Eq.~(\ref{eqA11})). We plotted $\Pi(\gamma,\alpha_{eq})$ in Figs.~\ref{fig5}(a,b). Figure~\ref{fig5}(a) demonstrates that the analytical equations describe $\Pi$ well. Thus, we shall use the analytically derived distribution $\Pi(\gamma,\alpha_{eq})$ (like the one shown in Fig.~\ref{fig5}(b)) to describe the trapping probability for the construction of a nonlocal operator acting on the full distribution function.

The coordinates of particle resonant interaction with the wave in the $(z, p_z)$ plane are defined by the resonant condition $\dot \phi+\dot\psi=0$ and the unperturbed particle trajectory ($\gamma=const$, $I_x=const$).
There is an additional condition for trapping, which can be formulated in a simplified form as a requirement of growth of the ratio of the wave force acting on the resonant particle over the combined mirror and inertial
forces \citep[see more accurate definition in, e.g.,][]{Neishtadt75, Neishtadt11:mmj}. Particles trapped at some $z_{trap}$ escape from the resonance at $z_{esc}$. The energy change is $\Delta \gamma_{trap}=\gamma_{R}(z_{esc})-\gamma_{R}(z_{trap})$, where energy $\gamma_R$ along the resonant trajectory is defined as (see Eq.~(\ref{eqA06}) in Appendix A and \citep[e.g.,][]{Artemyev18:jpp})
\begin{equation}
\gamma _R  = \left| {\varpi  \mp \frac{N}{{\sqrt {N^2  - 1} }}\sqrt {1 - 2\varepsilon _0 \varpi  + \varpi ^2 } } \right|
\label{eq03}
\end{equation}
In Eq.~(\ref{eq03}), $\varpi=\Omega_{ce}(z)/\omega$, $N=k(z)c/\omega$, and $\varepsilon_0=\gamma-\omega (\gamma^2-1)\sin^2\alpha_{eq}/2\Omega_{ce}(0)=const$. The sign in Eq.~(\ref{eq03}) is defined by the sign of the resonant velocity $\sim (1-\varpi/\gamma)/N$. For a simple cyclotron resonance with $\varpi/\gamma>1$ and negative resonant velocity $\sim (1-\varpi/\gamma)/N<0$, Eq.~(\ref{eq03}) should be used with a `$-$', while for the turning acceleration with $\varpi/\gamma<1$ and $(1-\varpi/\gamma)/N>0$ \citep[][]{Omura07}, Eq.~(\ref{eq03}) should be used with a `$+$'. To estimate $\Delta \gamma_{trap}$, we have to determine $z_{esc}$. The condition of particle escape from the resonance is determined by particle motion in the $(\zeta, \dot\zeta)$ phase plane, where $\zeta=\phi+\psi$. The computation of $z_{esc}$ and $\Delta \gamma_{trap}$ is described in Appendix B. The obtained analytical expression for
$\Delta \gamma_{trap}$ can be tested using numerical integration of the Hamiltonian equations (\ref{eq02a}). Figure~\ref{fig5}(c) compares numerical and analytical distributions $\Delta \gamma_{trap}(\gamma, \alpha_{eq})$. We
also indicate the energy gain due to the turning acceleration, when the direction of trapped particles motion changes and the particles gain significantly more energy than through an usual cyclotron resonance \citep[see
details in][]{Omura07}. This comparison shows that we can use the analytical distributions from Fig.~\ref{fig5}(d) to characterize electron trapping acceleration.

To complete the description, we need the energy change $\Delta \gamma_{scat}$. As nonlinear scattering is a local process, $\Delta \gamma_{scat}$ depends on wave properties
and background magnetic field characteristics at the resonant location in the $(z, p_z)$ plane: $\Delta \gamma_{scat}=-\omega S_{res}/2\pi$ where
\begin{linenomath}
\begin{equation}
S_{res}  = \sqrt {\frac{{8r}}{g}} \int\limits_{\zeta _ -  }^{\zeta _ +  } {\sqrt {a\left( {\sin \zeta _ +   - \sin \zeta } \right) - \left( {\zeta _ +   - \zeta } \right)} \;d\zeta }
\label{eq04}
\end{equation}
\end{linenomath}
$g=\omega^2(N^2-1)/m_e^2c^4\gamma_R$, and $r=r(z)$, $a=a(z)$ (see Appendix C and \citep[e.g.,][]{Artemyev14:pop}). The integration limits, $\zeta_{\pm}$, are defined in Eq.~(\ref{eqA10}) in Appendix C. The resonance location depends on the
initial $(\gamma, \alpha_{eq})$ and thus we can define $\Delta\gamma_{scat}(\gamma, \alpha_{eq})$. To check the analytical $\Delta\gamma_{scat}(\gamma, \alpha_{eq})$, we used the same approach as we used for $\Pi$. We bin $(\gamma, \alpha_{eq})$ space and for each pair of $\gamma, \alpha_{eq}$ values we numerically integrated $5\cdot 10^3$ trajectories of test particles. Then we calculated the energy change
after a single resonant interaction for each particle and averaged these changes over the non-trapped (scattered) particles. Figure~\ref{fig5}(e) shows $\Delta\gamma_{scat}(\gamma, \alpha_{eq})$ obtained from the numerical
integration of particle trajectories, and the analytically evaluated $\Delta\gamma_{scat}(\gamma, \alpha_{eq})$. This comparison confirms that we can use the analytical distributions $\Delta
\gamma_{scat}(\gamma,\alpha_{eq})$ from Fig.~\ref{fig5}(f) to describe the dynamics of an ensemble of charged particles.

\section{Kinetic equation\label{sec:kinetics}}

\subsection{Single wave\label{sec:kineticssingle}}
The nonlinear trapping and scattering of resonant electrons results in evolution of the electron distribution function $\Psi(\gamma,\alpha_{eq})$. To describe this evolution, we need a generalized kinetic equation
that includes effects of nonlinear scattering \citep[i.e., drift in energy space, see, e.g.,][]{Albert02} and trapping \citep[i.e., fast transport in energy space, e.g., see][]{Omura07}. Let us start with a system with one wave frequency, $\omega$. The particle energy in the wave reference frame is an integral of motion \citep[e.g.,][and references therein]{Summers98}. Using the
resonant condition $\dot\phi+\dot\psi=0$ (i.e., $kp_z-\gamma\omega+\Omega_{ce}=0$), this integral can be written as $m_ec^2\gamma-\omega I_x=const$ (alternatively,
$\gamma-\left(\omega/2\Omega_{ce}(0)\right)(\gamma^2-1)\sin^2\alpha_{eq}=const$ ). Thus, for systems with one $\omega$, the energy change directly defines the pitch-angle change
\begin{linenomath}
\begin{equation}
\Delta \alpha _{eq}  = \Delta \gamma \frac{{\Omega _{ce} (0) - \omega \gamma \sin ^2 \alpha _{eq} }}{{\omega \left( {\gamma ^2  - 1} \right)\sin \alpha _{eq} \cos \alpha _{eq} }}
\label{eq05}
\end{equation}
\end{linenomath}
This relation allows to compute $\Delta\alpha_{scat}$, $\Delta\alpha_{trap}$ once $\Delta\gamma_{scat}$, $\Delta\gamma_{trap}$ are known. The relation between $\gamma$ and $\alpha_{eq}$ can be used to examine the evolution of the 1D distribution function $\Psi(\gamma, \varepsilon_0)$ where $\varepsilon_0=\gamma-\left(\omega/2\Omega_{ce}(0)\right)(\gamma^2-1)\sin^2\alpha_{eq}$ is a constant parameter. The kinetic equation describing the evolution of a particle distribution has been derived in \citet{Artemyev18:jpp}. However, this kinetic equation does not allow a simple averaging over wave characteristics ($\omega$, $B_w$, etc.). Therefore, we follow the approach proposed by \citet[][]{Omura15}, and use $\Pi$, $\Delta\gamma_{trap}$, $\Delta\gamma_{scat}$ to derive the operator acting on $\Psi$. Note that in contrast to \citet[][]{Omura15},
we use here analytical expressions for $\Pi$, $\Delta\gamma_{trap}$, and $\Delta\gamma_{scat}$.

\begin{figure*}
\includegraphics[width=0.9\textwidth]{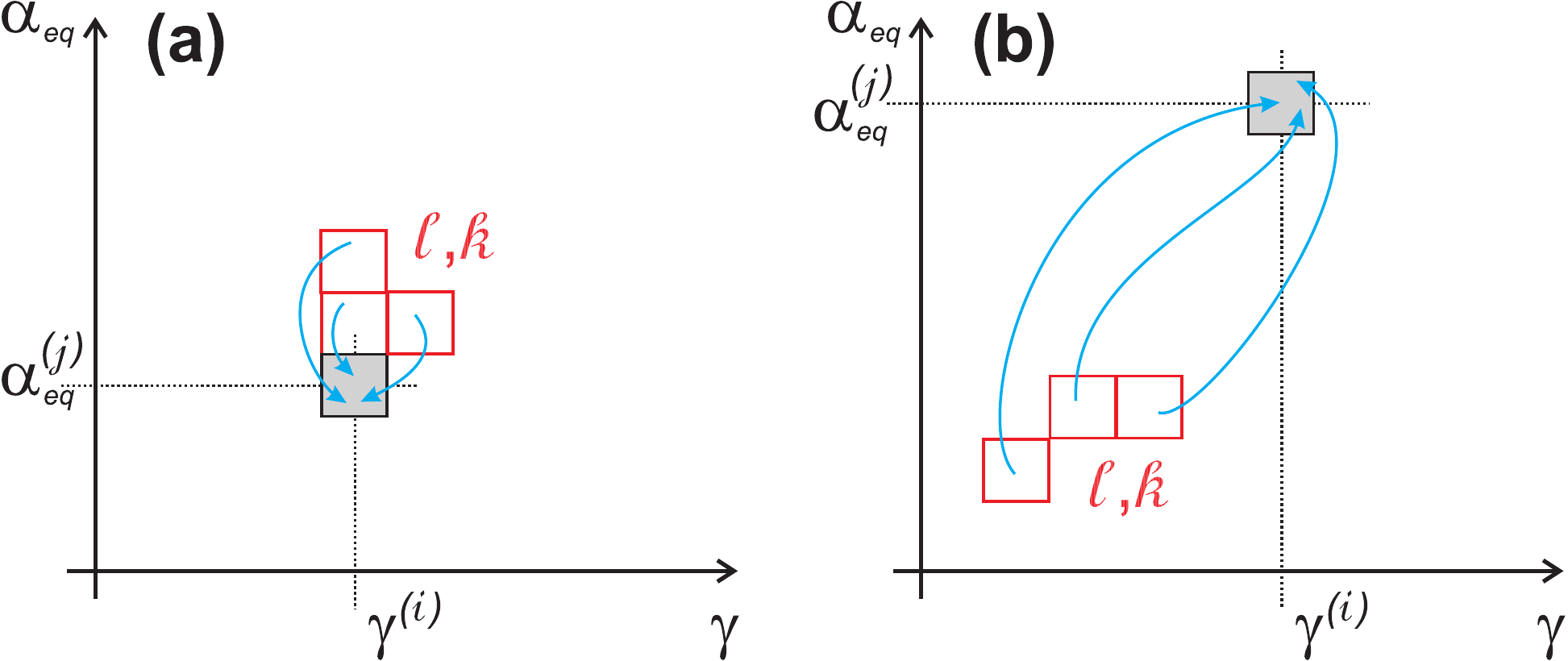}
\centering
\caption{Schematic view of operators describing (local) nonlinear scattering (a) and (nonlocal) trapping (b) in the energy/pitch-angle space. Red cells form the sets of $(k,l)$ indexes for scattered and trapped
particles.}
\label{fig6}
\end{figure*}

We bin the $(\gamma,\alpha_{eq})$ space as $\gamma^{(i)}=1+\Delta_\gamma i$, $\alpha_{eq}^{(j)}=\Delta_\alpha j$ where $i,j=0...M$ and $\Delta_\alpha=\pi/M$, $\Delta\gamma=(\gamma_{\max}-1)/M$. Then for each bin
$(\gamma^{(i)},\alpha_{eq}^{(j)})$, we define $\Pi^{ij}$, $\Delta\gamma_{scat}^{ij}$, $\Delta\gamma_{trap}^{ij}$, and introduce two quantities: $s_{mn}^{kl}(W)$ and  $p_{mn}^{kl}(W)$ denoting the probabilities for a
particle to move from the state $(\gamma^{(k)},\alpha_{eq}^{(l)})$ to the state $(\gamma^{(m)},\alpha_{eq}^{(n)})$ due to a single nonlinear scattering ($s_{mn}^{kl}$) and trapping ($p_{mn}^{kl}$). The letter $W$
indicates all the relevant wave-packet characteristics, most importantly its amplitude $B_w$ and frequency $\omega$. For each packet (fixed $B_w$, $\omega$), one can view all $s^{kl}_{mn}(W)$, $p^{kl}_{mn}(W)$ as the
elements of a big 4D matrix that defines the phase space transport due to scattering and trapping. Both $s_{mn}^{kl}(W)$ and  $p_{mn}^{kl}(W)$ are obtained using $\Pi^{ij}$, $\Delta\gamma_{trap}^{ij}$, and
$\Delta\gamma_{scat}^{ij}$:
\begin{linenomath}
\begin{equation}
\begin{array}{l}
 s_{mn}^{kl}  = \left\{ \begin{array}{l}
 1 - \Pi ^{kl} ,\quad \begin{array}{*{20}c}
   {\gamma ^{(k)}  + \Delta \gamma _{scat}^{kl}  \in \left[ {\gamma ^{(m)}  - \textstyle{{\Delta _\gamma  } \over 2}  ,\;\gamma ^{(m)}  + \textstyle{{\Delta _\gamma  } \over 2}  } \right]}  \\
   {\alpha _{eq}^{(l)}  + \Delta \alpha _{eq,scat}^{kl}  \in \left[ {\alpha _{eq}^{(n)}  - {\textstyle{{\Delta _\alpha} \over 2}}  ,\;\alpha _{eq}^{(n)}  + {\textstyle{{\Delta _\alpha} \over 2}} } \right]}  \\
\end{array} \\
 0,\quad {\rm otherwise} \\
 \end{array} \right. \\
\\ \label{eq06}
 p_{mn}^{kl}  = \left\{ \begin{array}{l}
 \Pi ^{kl} ,\quad \begin{array}{*{20}c}
   {\gamma ^{(k)}  + \Delta \gamma _{trap}^{kl}  \in \left[ {\gamma ^{(m)}  - \textstyle{{\Delta _\gamma  } \over 2}  ,\;\gamma ^{(m)}  + \textstyle{{\Delta _\gamma  } \over 2}  } \right]}  \\
   {\alpha _{eq}^{(l)}  + \Delta \alpha _{eq,trap}^{kl}  \in \left[ {\alpha _{eq}^{(n)}  - {\textstyle{{\Delta _\alpha} \over 2}}  ,\;\alpha _{eq}^{(n)}  + {\textstyle{{\Delta _\alpha} \over 2}}  } \right]}  \\
\end{array}\quad  \\
 0,\quad {\rm otherwise} \\
 \end{array} \right. \\
 \end{array}
\end{equation}
\end{linenomath}
Consider a state $(\gamma^{(i)},\alpha_{eq}^{(j)})$. During each bounce period $\tau_{ij}=\tau(\gamma^{(i)},\alpha_{eq}^{(j)})$, a particle undergoes $n_{ij}=n(\gamma^{i},\alpha_{eq}^{j})$ resonant interactions.
Combining nonlinear scattering and trapping, we obtain a joint evolution equation of the electron distribution in the case of a single wave:
\begin{linenomath}
\begin{equation}
\frac{\partial \Psi_{ij}}{\partial t}= - \frac{n_{ij}}{\tau_{ij}} \Psi_{ij} + \sum_{kl} \frac{n_{kl}}{\tau_{kl}} s^{kl}_{ij}(W) \Psi_{kl}  + \sum_{kl} \frac{n_{kl}}{\tau_{kl}} p^{kl}_{ij}(W) \Psi_{kl}
\label{eq07}
\end{equation}
\end{linenomath}
where $\Psi_{ij}=\Psi(\gamma^{(i)},\alpha_{eq}^{(j)})$. The summation is performed over a set of indices $(k,l)$ for which $s^{kl}_{ij}$ or $p^{kl}_{ij}$ is non-zero -- in other words, over all the states
$(\gamma^{(k)},\alpha_{eq}^{(l)})$ from which particles can come to the state $(\gamma^{(i)},\alpha_{eq}^{(j)})$ after a single nonlinear scattering or trapping, as shown in Fig.~\ref{fig6}. The negative term
proportional to $\Psi_{ij}$ describes the removal of particles from the state $(\gamma^{(i)},\alpha_{eq}^{(j)})$, while the terms with the sum describe the incoming flux of particles. There is a noticeable difference between $s^{kl}_{mn}(W)$ and $p^{kl}_{mn}(W)$: while the non-null elements of $s^{kl}_{mn}(W)$ correspond to nearby cells
(around $k=m, l=n$), the elements of $p^{kl}_{mn}(W)$ are more remote, see Fig.~\ref{fig6}.

Equation (\ref{eq07}) can be rewritten as
\begin{eqnarray}
\frac{\partial \Psi_{ij}}{\partial t}&=& - \frac{n_{ij}}{\tau_{ij}} \Psi_{ij} + \sum_{kl} R^{kl}_{ij}(W) \Psi_{kl} \nonumber \\
R^{kl}_{ij}(W) &=& \frac{n_{kl}}{\tau_{kl}} \left( s^{kl}_{ij}(W) + p^{kl}_{ij}(W) \right)
\label{eq08}
\end{eqnarray}
Elements of $R^{kl}_{ij}(W)$ depend on wave characteristics. Note that Eq.(\ref{eq08}) is linear with respect to $\Psi_{ij}$ and all the nonlinearity is included in $R^{kl}_{ij}(W)$. Equation (\ref{eq08}) fully defines the evolution of $\Psi_{ij}$ in the presence of a single wave (i.e., for a given $W$). To generalize the results obtained  to an ensemble of waves, we need to average Eq.~(\ref{eq08}) weighted by the probability (occurrence rate) of the different wave parameters (different sets of $W$).

\subsection{Multiple waves\label{sec:kineticsmany}}

\begin{figure*}
\includegraphics[width=0.9\textwidth]{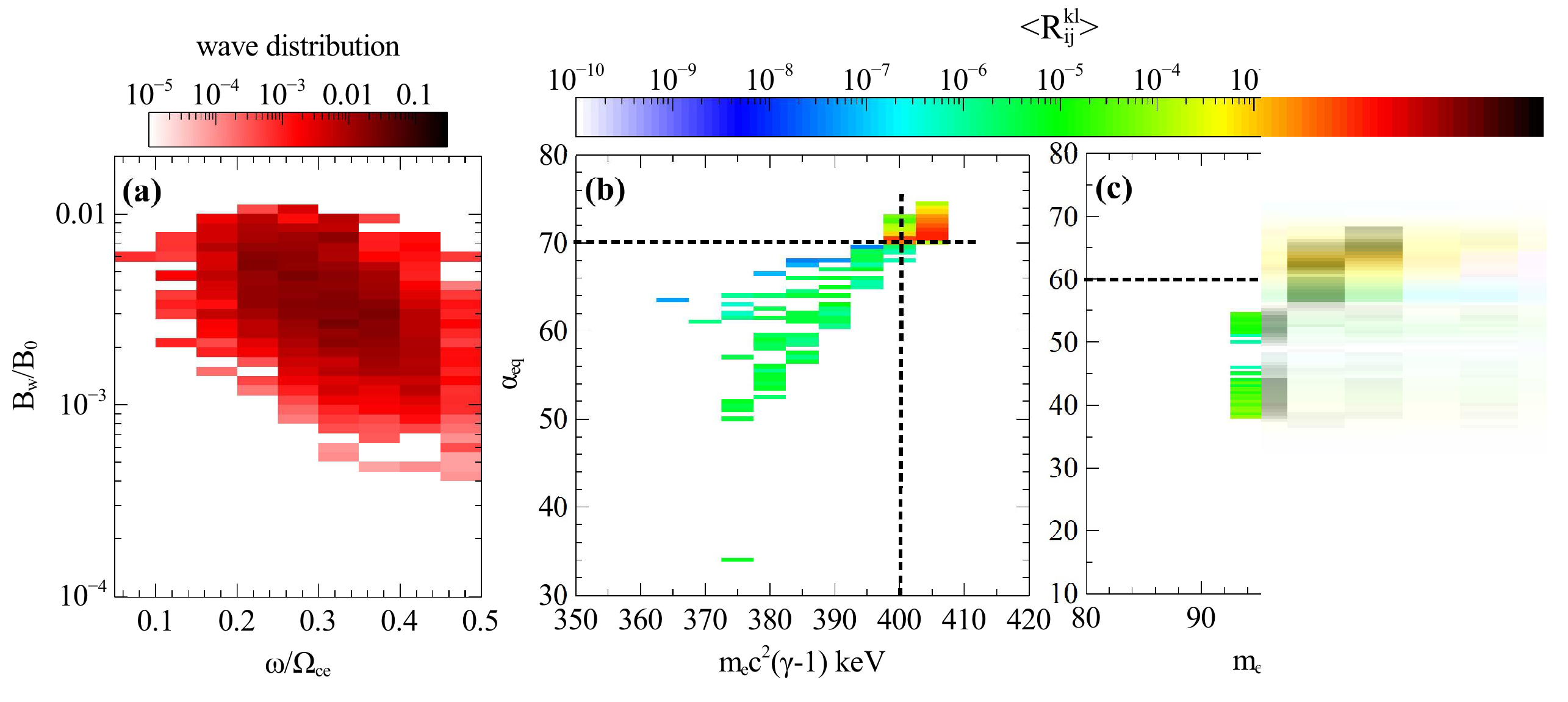}
\centering
\caption{(a) A model wave distribution in the ($B_w$, $\omega/\Omega_{ce}$) space (taken from Fig.~\ref{fig3} for $L*<6.6$). (b) and (c): Two functions $\langle R^{kl}_{ij}\rangle$ plotted for two {\it target} cells:
$(m_ec^2(\gamma^{(i)}-1) = 400$keV, $\alpha_{eq}^{(j)} = 70^\circ)$ (b) and $(m_ec^2(\gamma^{(i)}-1) = 100$keV, $\alpha_{eq}^{(j)} = 60^\circ)$ (c). Black lines in panels (b), (c) cross at the {\it target} cells. }
\label{fig7}
\end{figure*}

Consider an ensemble of waves, where $\rho(W)$ defines the normalized statistical weight of a certain set of parameters (e.g., each of the two probability distributions in Fig.~\ref{fig3}  could result in an empirical determination of $\rho(W)$ as function of wave amplitude $B_w$ and frequency $\omega$). Here, however, the distribution $\rho(W)$ is normalized so that $\int{\rho(W)dW}=T_{int}/T_{tot}$, where $T_{tot}$ is the total time interval of spacecraft wave measurements and $T_{int}$
is the cumulative time interval of observations of intense ($\bw>2$) and long chorus wave-packets. Instead of Eq.~(\ref{eq08}), we obtain
\begin{linenomath}
\begin{equation}
\frac{\partial \Psi_{ij}}{\partial t}= \int_W \left( - \frac{n_{ij}}{\tau_{ij}} \Psi_{ij} + \sum_{kl} R^{kl}_{ij}(W) \Psi_{kl} \right) \rho(W) d W
\label{eq09}
\end{equation}
\end{linenomath}
As $\Psi_{ij}$ does not explicitly depend on the wave properties $W$, it can be taken out of the integral. Moreover, in the first approximation (for relatively narrow-band parallel chorus waves interacting with
electrons through the single fundamental cyclotron resonance), $n_{ij}$ and $\tau_{ij}$ are roughly independent of $W$. Thus we get
\begin{linenomath}
\begin{eqnarray}
\frac{\partial \Psi_{ij}}{\partial t} &=& - \frac{n_{ij}}{\tau_{ij}} \frac{T_{int}}{T_{tot}}\Psi _{ij} + \left( \int_W \sum_{kl} R^{kl}_{ij}(W) \rho(W) d W \right) \Psi _{kl} \nonumber\\
&=& - \frac{n_{ij}}{\tau_{ij}} \frac{T_{int}}{T_{tot}}\Psi _{ij} + \sum_{kl} \langle R^{kl}_{ij}\rangle \Psi _{kl}
\label{eq10}
\end{eqnarray}
\end{linenomath}
where $\langle R^{kl}_{ij}\rangle = \int_W R^{kl}_{ij}(W) \rho(W) d W$. Typical plots of $\rho(W)$ and $\langle R^{kl}_{ij}\rangle$ are shown in Fig.~\ref{fig7}. For the first-order cyclotron resonance, nonlinear scattering results in energy/pitch-angle decrease and trapping results in energy/pitch-angle increase. Therefore, for a given state $(\gamma^{(i)},\alpha_{eq}^{(j)})$, the matrix $\langle R^{kl}_{ij}\rangle$ contains two groups of non-zero elements: (i) elements at energies/pitch-angles slightly larger than the {\it target} cell describe the efficiency of nonlinear scattering; (ii) elements located at energies/pitch-angles quite smaller than the {\it target} cell describe the efficiency of trapping.

\section{Evolution of the electron distribution \label{sec:distribution}}
Equation~(\ref{eq10}) describes the long-term evolution of the electron distribution function due to nonlinear wave-particle interaction. To demonstrate how this theoretical approach works for realistic waves, we
parameterize the distribution of prolonged and intense chorus waves from Fig.~\ref{fig3}b (at $L^*<6.6$) and substitute it as $\rho(W)$ into Eq.~(\ref{eq10}). We use the normalized time $\tau=t(c/R_E)(T_{int}/T_{tot})$ and thus the $\rho$ distribution is normalized to unity ($\int\rho dW=1$). We take $T_{int}/T_{tot}\approx 10^{-3}$ during active periods with $AE\geq 500$ nT, in agreement with the chorus wave statistics showing that long wave-packets with $\beta>50$ represent $\sim 2$\% of intense wave observations, which themselves are present during $\sim 10$\% of lower-band chorus wave measurements performed during active periods between $20$ MLT and $10$ MLT -- the MLT range of intense chorus emissions, representing 58\% of the full MLT range \citep[see][]{Zhang18:jgr:intensewaves, Mourenas18:jgr}. Therefore, normalized time $\tau=1$ corresponds to $\sim 21$ seconds in real time.

\begin{figure*}
\includegraphics[width=0.9\textwidth]{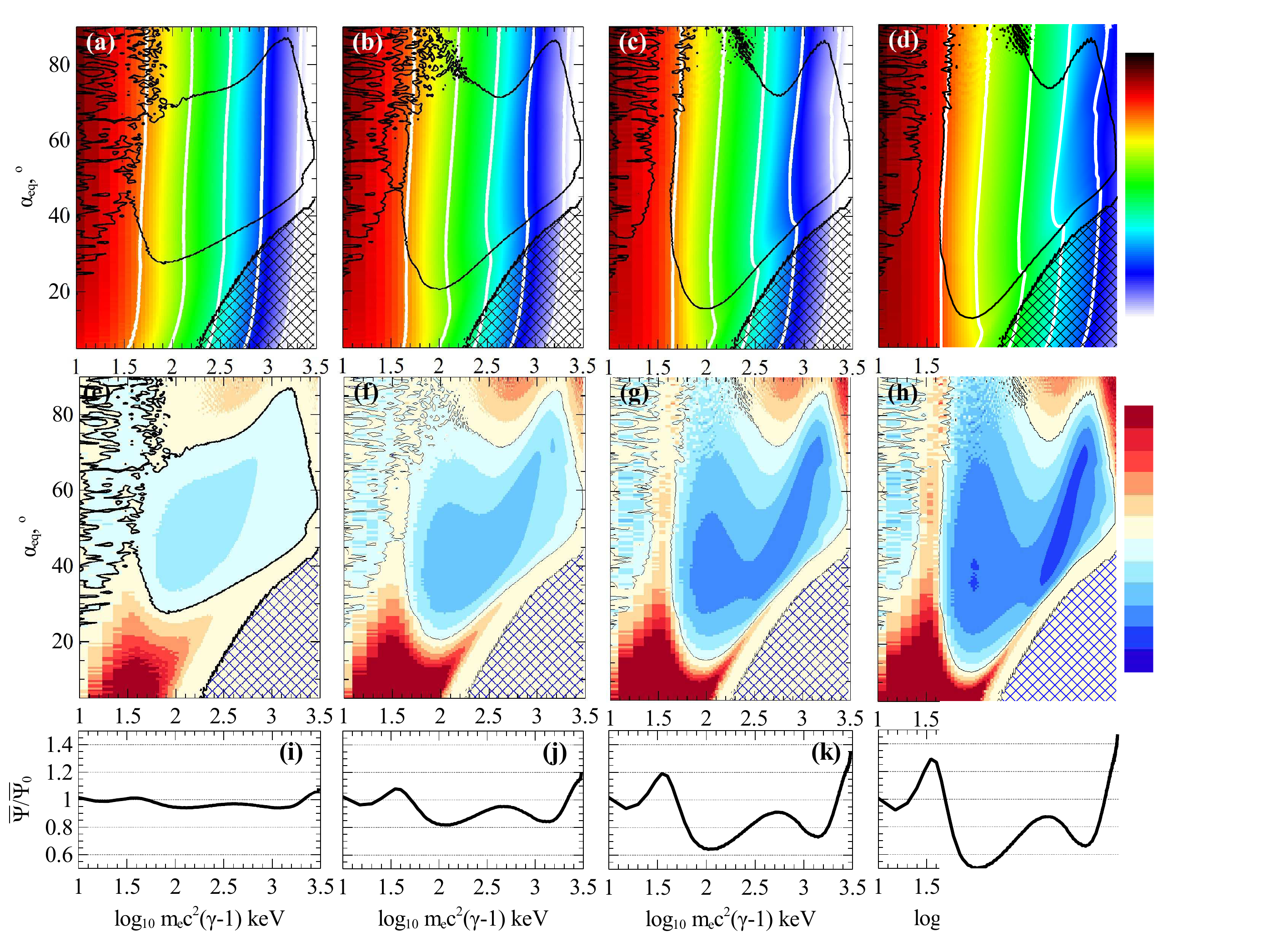}
\centering
\caption{Evolution of the electron distribution function due to nonlinear interactions with intense chorus waves. Panels (a)-(d) show the solution of Eq.~(\ref{eq10}) for an initial distribution
$\Psi_0=(1+(\gamma-1)/c_{\kappa})^{-\kappa-1}\sin\alpha_{eq}$ with $c_{\kappa}=(\kappa-3/2)/50$ and $\kappa = 4$. The results of numerical calculations are displayed at times $\tau=10$ (a), $\tau=30$ (b), $\tau=60$ (c),
and $\tau=100$ (d). Panels (e)-(h) show the ratio of distributions from panels (a)-(d) over the initial distribution. Panels (i)-(l) show the integrated distributions $\bar \Psi=\int\Psi\sin\alpha_{eq}d\alpha_{eq}$
normalized to the initial integrated distribution $\bar \Psi_0$. Shadowed domains in panels (a)-(h) show the parameter domains of non-resonant particles. Solid black curves in panels (a)-(h) correspond to $\Psi/\Psi_0 =
1$.}
\label{fig8}
\end{figure*}

Both nonlinear scattering and trapping have a pronounced effect on $\Psi(\gamma,\alpha_{eq})$ and $\bar\Psi(\gamma)$. Figure~\ref{fig8} shows $\Psi(\gamma,\alpha_{eq})$ and $\bar\Psi(\gamma)=\int_0^{\pi/2}{\Psi(\gamma,\alpha_{eq})\sin\alpha_{eq} d\alpha_{eq}}$ at $\tau=10$ ($\sim 4$ minutes), $\tau=30$ ($\sim 10$ minutes), $\tau=60$ ($\sim 21$ minutes), and $\tau=100$ ($\sim 35$ minutes) of the nonlinear wave-particle interaction. The main feature of the distribution displayed in Fig.~\ref{fig8} is a heavily reduced middle
domain at $\alpha_{eq}\in [30^{\circ}, 70^{\circ}]$ and energies $\in [10^2, 10^3]$ keV). Electron interaction with an ensemble of intense chorus waves leads to an efficient nonlinear scattering toward smaller
energies/pitch-angles ($\Psi$ increases at energies $\sim 30$ keV below the domain of reduced phase space density) and trapping acceleration to larger energies ($\Psi$ increases at energies $\sim 3$ MeV above the domain
of reduced phase space density). Both the decelerated low-energy electron population and the accelerated high-energy population (see peaks in  the energy spectra $\bar\Psi(\gamma)$ increasing around 30 keV and 3 MeV in
the bottom panels of Fig.~\ref{fig8}) are formed after a relatively short time ($\sim 35$ minutes) when considering the time-averaged characteristics of intense chorus waves from Fig.~\ref{fig3}b -- in spite of the presence of 30-40 \% waves with with $B_w/B_0<0.0035$ generally not sufficiently intense to produce NL effects. Moreover, the occurrence rate of long and intense chorus wave-packets can become 3-5 times higher than  time-averaged level used. This can occur over short time intervals ($\sim 10$ minutes) corresponding to plasma injections during dipolarization events at $L^*\sim 5$ \citep{Zhang18:jgr:intensewaves}. In principle, this could result in an even faster evolution of $\Psi$ at low electron energy $<100$ keV (corresponding to an azimuthal drift period $>1$ hour), but barely accelerates the evolution of $\Psi$ at $0.4-3$ MeV because the azimuthal drift period of such electrons is
5-25 minutes, which implies that the particles encounter such rare events (localized in MLT) during only a small fraction of one azimuthal drift period. However, we caution that various other effects can modify the
efficiency of nonlinear interaction and that they should be carefully taken into account to estimate the actual time scale of the evolution of the electron distribution. These  will be discussed in
details in section \ref{sec:discussion}.

\begin{figure*}
\includegraphics[width=0.9\textwidth]{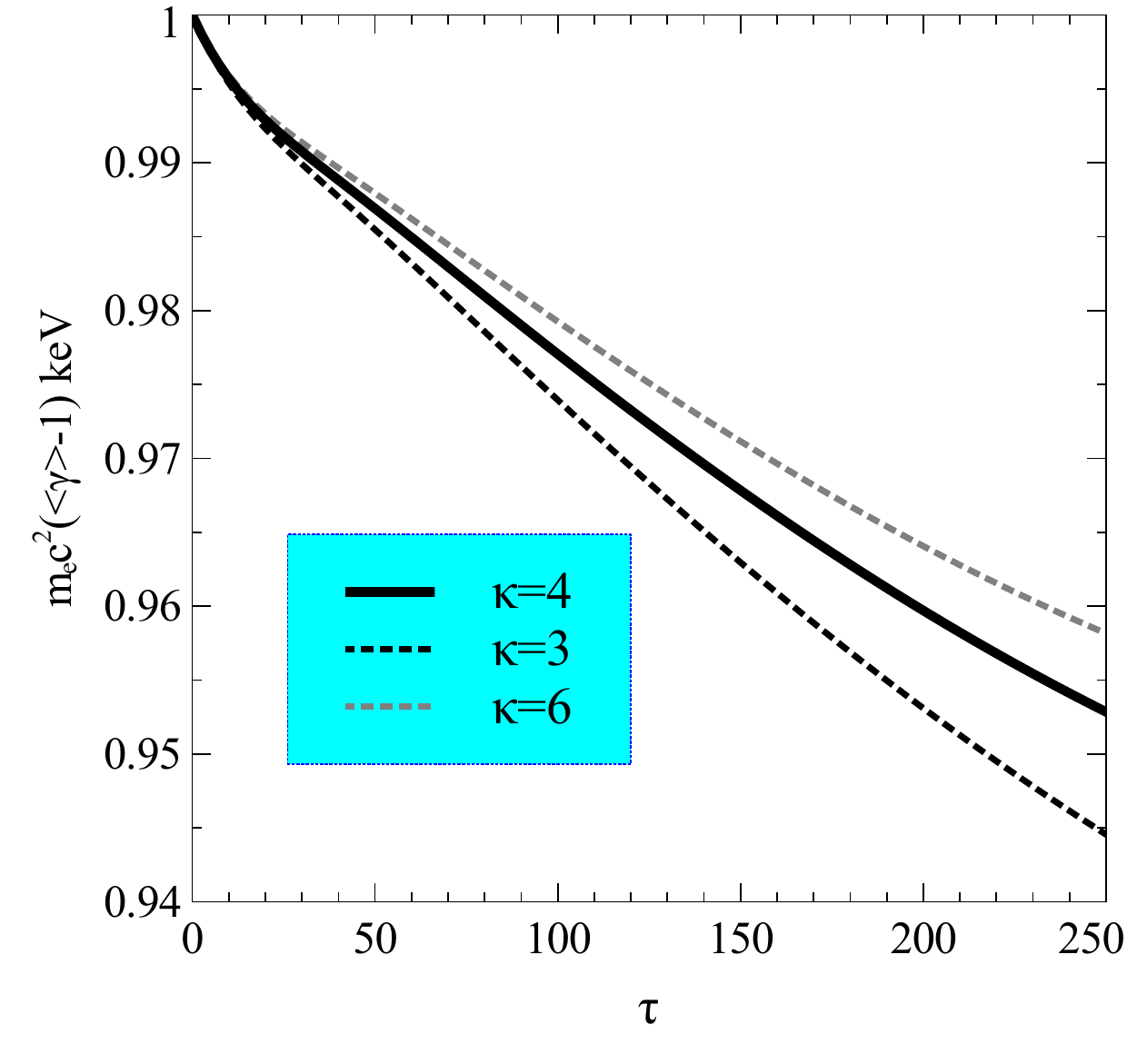}
\centering
\caption{Left panel: evolution of the total particle energy for initial distributions $\Psi_0=(1+(\gamma-1)/c_{\kappa})^{-\kappa-1}\sin\alpha_{eq}$ with $c_{\kappa}=(\kappa-3/2)/50$ and three values of $\kappa$. Right panel: Evolution of total particle energy for initial distributions $\Psi_0=(1+(\gamma-1)/c_{\kappa})^{-\kappa-1}G(\alpha_{eq})$ with four different $G(\alpha_{eq})$ functions; $c_{\kappa}=(\kappa-3/2)/50$ and
$\kappa=4$.}
\label{fig9}
\end{figure*}

Figure~\ref{fig8} demonstrates the presence of both acceleration and deceleration of resonant electrons: nonlinear scattering results in particle drift to smaller energies and trapping transports particles to higher
energies. These two processes should be in fine balance due to the strong relation linking the probability of trapping and the scattering drift rate \citep[see, e.g.,][]{Solovev&Shkliar86, Artemyev18:jpp}. The analytical theory predicts that the total energy change of resonant electrons should not be large \citep{Shklyar11:angeo}, because any significant energy change necessitates some energy source or sink. Thus if the total energy varies, either the assumption of roughly constant wave intensity is not satisfied or the initial electron distribution was not chosen properly. To check the prediction of small energy change we plotted the average particle energy $m_ec^2\langle\gamma-1\rangle$ as a function of time for $\Psi$ from Fig.~ \ref{fig8}. A weak but still significant (several percent) decrease of $m_ec^2\langle\gamma-1\rangle$, indicates that the initial distribution $\Psi$ may not have been chosen fully properly, Fig.~\ref{fig9}(left panel). However, the slope of the initial energy distribution (the $\kappa$ parameter) does not significantly influence the rate of decrease.

\section{Discussion \label{sec:discussion}}
In this study, we focused on the electron nonlinear resonant interactions with long and intense chorus wave-packets which can support a rapid evolution of the electron distribution via trapping and nonlinear scattering.
When combined with the previous investigations of electron diffusion by low-intensity waves \citep[see][and references therein]{bookSchulz&anzerotti74,bookLyons&Williams} and intense short wave-packets
\citep[see][]{Mourenas18:jgr}, the present investigation complements the set of theoretical tools needed to fully describe electron dynamics driven by realistic whistler-mode waves in the radiation belts. Here, we
discuss several important questions arising from a careful analysis of solutions of Eq.~(\ref{eq10}).

\subsection{Time scale of electron distribution evolution}
Figure~\ref{fig8} shows that the typical time scale of the evolution of the electron distribution via nonlinear trapping and scattering by long and intense chorus wave-packets can be about half an hour, i.e., rather short compared with $\sim 4-10$ hours typically for quasi-linear diffusion by the bulk of lower-intensity chorus waves \citep[e.g.,][]{Mourenas14, Li16:jgr, Ma17:jgr, Yang18}. Although such fast electron flux variations have
been occasionally observed \citep[e.g.,][]{Agapitov15:grl:acceleration,Foster17}, they are rather unusual. Therefore, we discuss below the possible effects that can slow down the nonlinear
evolution of the electron distribution.

First, during low geomagnetic activity the occurrence rate of intense chorus waves is reduced \citep[see the dependence of intense wave occurrence rate on the $AE$ index in][]{Zhang18:jgr:intensewaves}. Accordingly, the characteristic time scale of electron acceleration would be considerably increased as compared with $AE\geq 500$ nT. Moreover, even during geomagnetically active periods,
intense long wave-packets are observed only sporadically. Time intervals when such long wave-packets are present mainly correspond to intervals ($\sim 10$ minutes) of significant plasma injections that provide a free energy source for strong wave generation \citep[e.g.,][]{Tao11, Demekhov17, Zhang18:jgr:intensewaves}. Such realistic chorus wave-packets are limited in time and space \citep[spatial scales of the source region of chorus waves are about $\sim 500-1000$ km, see][]{Agapitov17:grl}. Thus, they have a smaller (and slower) effect on the electron distribution than those in Fig.~\ref{fig8}, where we considered very long wave-packets present all the time with their time-averaged occurrence rate. Quasi-linear diffusion coefficients can be directly calculated by time-averaging the wave intensity. To obtain the nonlinear resonant
operator in Eq.~(\ref{eq10}) we can average operator over a distribution of long intense wave-packet characteristics and multiple it by the percentage of time when such wave-packets are observed. However, nonlinear effects described by this operator may be significantly smoothed and slowed down (compared with results of Fig.~\ref{fig8}), if short periods of really intense wave-packet observations are mixed with long periods of moderately intense waves. This may partly explain why the effects of the nonlinear wave-particle interaction (rapid electron energization/scattering) can be observed mostly within short periods of intense waves. These effects are barely noticeable in the long-term dynamics of the outer radiation belt.

Second, there is an important additional factor that can reduce the efficiency of the nonlinear wave-particle interaction, slowing down the global evolution of the electron distribution. Even long wave-packets may not
always sustain a prolonged nonlinear interaction, because of a possible destruction of the nonlinear resonance by adverse effects \citep[caused by additional resonant sidebands of lower amplitude, or non-resonant weaker waves, see][]{Nunn86, Shklyar&Zimbardo14, Artemyev15:pop:stability}. The length (and/or amplitude) of long wave-packets should also probably decrease as they propagate toward higher
latitudes \citep{Tsurutani11}. This effect reduces the rate of trapping acceleration \citep[e.g.,][]{Artemyev12:pop:nondiffusion}. Intense chorus wave statistics presented in Fig.~\ref{fig3} were obtained at low latitudes, inside or near the wave source region. Moreover, Fig.~\ref{fig3}(a) indicates that important modulations of the wave intensity are generally present inside long wave-packets,
which can result in a further reduction of the efficiency of nonlinear wave-particle interaction \citep{Tao13, Artemyev15:pop:stability}. Only $\sim 15-20$\% of intense long wave-packets (with $\beta>50$) contain at least
$\sim 50$ wave-periods with a wave amplitude larger than the half of the peak wave-packet amplitude, i.e., most of the long wave-packets are rather localized, with a narrow peak of wave intensity and much less intense
{\it tails}. Based on this last effect alone, the typical time scale of nonlinear evolution of the electron distribution can be increased by a factor $\sim 5-6$ as compared with Fig.\ref{fig8}, reaching already $3-4$ hours,
similar to quasi-linear time scales.

\subsection{Nonlinear wave-particle interaction vs. quasi-linear models}
Equation (\ref{eq10}) describes changes of the electron distribution $\Psi$ due to the nonlinear wave-particle interaction. This equation has the form of a classical evolution equation
\citep[e.g.,][]{book:VanKampen03}: the time derivative $\partial \Psi/\partial t$ is equal to an operator $\hat L$ acting on $\Psi$: $\partial \Psi/\partial t=\hat L \Psi$. The same type of
equation describes quasi-linear particle diffusion by an ensemble of low-intensity waves, $\partial \Psi/\partial t=\hat D \Psi$ \citep[][]{bookSchulz&anzerotti74}, where the diffusion operator $\hat D \sim
\partial^2/\partial\gamma^2, \partial^2/\partial \alpha_{eq}^2$ is composed of quasi-linear diffusion rates that depend on the wave intensity and dispersion \citep[see examples in][]{Glauert&Horne05, Albert08,
Shprits&Ni09}. Therefore, Eq.(\ref{eq10}) can be readily incorporated into existing numerical codes solving diffusion equations. The resultant combined equation would include three operators $\partial \Psi/\partial
t= T_{D}\hat D\Psi+T_{L}\hat L \Psi+T_{K}\hat K \Psi $ where coefficients $T_{D}$, $T_{L}$, and $T_{K}$ define the relative time intervals of spacecraft observations of low-intensity ($\bw<2$, quasi-linear regime), high-intensity long wave-packets ($\bw>2$, $\beta^*>50$, nonlinear regime), and high-intensity short wave-packets ($\bw>2$, $\beta^*<50$, nonlinear regime) waves. The operator $\hat K$ describing the combined action of nonlinear trapping and scattering by short wave-packets has been derived by \citet{Mourenas18:jgr} and amounts to  a simple drift term $\hat
K=-W_\gamma\partial\Psi/\partial \gamma-W_\alpha\partial\Psi/\partial \alpha_{eq}$. For existing Fokker-Planck numerical codes of the radiation belt dynamics, the operator $\hat L$ can be considered as a source/loss
operator \citep[similar to the $\tau$-operator describing particle losses $\partial \Psi/\partial t \sim -\Psi/\tau$ -- e.g., see][]{Horne05JGR, Balikhin12, Mourenas14}.

\subsection{Energy conservation}
The evolution of the electron distribution function due to the nonlinear wave-particle interaction includes a rapid change of the total electron energy (see Fig.~\ref{fig9}(left panel)). However, the energy of a whistler-mode wave is usually not sufficiently large to provide such a deceleration/acceleration \citep[see discussion and estimates in][]{Shklyar11:angeo} and thus scattered (decelerated) electrons are the only available energy source for trapped electron acceleration. This balance imposes certain constraints on the initial electron distribution function for models describing nonlinear resonances. This initial distribution should be consistent with conservation of the total energy for solutions of Eq.~(\ref{eq10}). Figure~\ref{fig9}(left panel) shows that one cannot easily reduce the rate of the total energy decrease simply by choosing a different electron energy distribution. We further checked the total particle energy variation for different initial pitch-angle distributions. Figure~ \ref{fig9}(right panel) shows that an initial distribution with transverse anisotropy ($\Psi \sim \sin\alpha_{eq}$) immediately loses energy, whereas a field-aligned anisotropic
distribution ($\Psi \sim \cos\alpha_{eq}$) first gains energy before losing it. We also checked more specific distributions (butterfly $\Psi\sim \sin2\alpha_{eq}$ sometime observed in the radiation belts \citep[see
statistics in][]{Asnes05} and a distribution with field-aligned beams $\Psi\sim \sin\alpha_{eq}\cos^4\alpha_{eq}$ observed during particle injections \citep[e.g.,][]{Mozer16}). Both show energy
increase during the first stage of the evolution. This suggests that for a $\Psi$ to result in total energy conservation it should be more isotropic (i.e., something between transversely anisotropic
$\sim\sin\alpha_{eq}$ and field-aligned anisotropy $\sim \cos\alpha_{eq}$). Moreover, under realistic conditions, particle anisotropy is energy dependent, and can be different for $<100$ keV particles and $\sim 1$ MeV
particles. Therefore, further investigations of particle distributions supporting the assumption of total energy conservation are needed for an accurate modelling the nonlinear wave-particle interaction \citep[see, also,][for
details on the relation between trapped and scattered electron populations in a system with small variation of wave intensity]{Shklyar17}.

\section{Conclusions}
We investigated the nonlinear resonant electron interaction with intense whistler-mode waves in the outer radiation belt. Combining THEMIS and Van Allen Probe observations of long and intense chorus wave-packets with the
analytical theory of the resonant wave-particle interaction, we constructed a generalized kinetic equation that describes the evolution of the electron distribution function. The main conclusions of this study are:
\begin{itemize}
\item The observed occurrence rate and characteristics (amplitude, frequency) of intense parallel chorus waves may provide a very rapid (over tens of minutes) electron acceleration in the outer radiation belt. But such
    a rapid acceleration should be observed only sporadically and locally, during time intervals of plasma injection and strong wave generation. Otherwise, the combination of very small time-averaged occurrence rate of long and  intense chorus wave-packets and various other  effects moderating their efficacy, lead to acceleration, occurring over typical times scales $>3-4$ hours similar to, or
    possibly larger than, quasi-linear diffusion time scales.
\item Nonlinear trapping and scattering cause an extremely intense energy exchange between different particle populations. In the absence of a fine balance between these different particle populations, particle
    acceleration (deceleration) would immediately result in intense wave damping (growth). Thus, the observations of intense chorus wave emissions during many periods of resonant interaction suggest that these waves may be
    accompanied by particular (balanced) distribution of electrons, evolving with almost no variation of their total energy.
\item Nonlinear effects of the wave-particle interaction can be included into modern codes of radiation belt dynamics, but to exclude unrealistically large particle acceleration/deceleration, such effects should be
    considered only with properly chosen initial particle distributions.
\end{itemize}

\section*{Appendix A: probability of trapping}
In this Appendix we derive the probability of electron trapping into resonance \citep[see details in, e.g.,][]{Artemyev15:pop:probability, Artemyev18:jpp}. We start with Hamiltonian (\ref{eq02}) and
follow the proposed in \citet{Neishtadt99}. First, we introduce the momentum $I$ conjugate to phase $\zeta=\phi+\psi$. We use generating function
$S_1=P_zz+I(\phi+\psi)+\tilde {I}_{x}\psi$ where $P_z = p_z-kI$ and $\tilde{I_x}=I_x-I$ are new momenta conjugate to unchanged coordinates $z$ and $\psi$. The new Hamiltonian $F=H+\partial S_1/\partial t$ is
\begin{linenomath}
\begin{eqnarray}
F &=&  - \omega I + m_e c^2 \gamma  + U_w (z,\tilde I_x  + I)\sin \zeta\nonumber\\
\gamma  &=& \sqrt {1 + \frac{{\left( {P_z  + kI} \right)^2 }}{{m_e^2 c^2 }} + \frac{{2\Omega _{ce} }}{{m_e c^2 }}\left( {\tilde I_x  + I} \right)}
\label{eqA01}
\end{eqnarray}
\end{linenomath}
Hamiltonian (\ref{eqA01}) does not depend on $\psi$, and thus $\tilde{I}_x$ is constant (we chose $\tilde{I}_x=0$, i.e., $I=I_x$ at the initial time). Moreover, Hamiltonian (\ref{eqA01}) does not explicitly depend on time either ($\zeta$ is a new variable), and thus $F=const$. This constant can be written as $F/m_ec^2=\varepsilon_0=\gamma-\omega (\gamma^2-1)\sin^2\alpha_{eq}/2\Omega_{ce}(0)$, where we took into account that
$I_x=(1/2)(\gamma^2-1)\sin^2\alpha_{eq}/\Omega_{ce}(0)$ and $I=I_{x}$. The resonance condition ($\dot\zeta=0$) for Hamiltonian (\ref{eqA01}) is
\begin{linenomath}
\begin{equation}
\frac{{\partial F}}{{\partial I}} =  - \omega  + \frac{{k\left( {P_z  + kI} \right)}}{{m_e \gamma }} + \frac{{\Omega _{ce} }}{\gamma } = 0
\label{eqA02}
\end{equation}
\end{linenomath}
Combination of Eq.~(\ref{eqA01}) and Eq.~(\ref{eqA02}) gives for $I=I_R(z,P_z)$ at the resonance:
\begin{linenomath}
\begin{equation}
\frac{{\omega I_R }}{{m_e c^2 }} =  - \frac{\varpi }{{N^2 }} - \frac{{P_z }}{{Nm_e c}} + \frac{1}{{N\sqrt {N^2  - 1} }}\sqrt {1 - \frac{{\varpi ^2 }}{{N^2 }} - 2\frac{\varpi }{N}\frac{{P_z }}{{m_e c}}}
\label{eqA04}
\end{equation}
\end{linenomath}
where $N=k(z)c/\omega$, $\varpi=\Omega_{ce}(z)/\omega$. Substituting Eq.~(\ref{eqA04}) into $\gamma$ from Eq.~(\ref{eqA01}), we obtain the resonant energy:
\begin{linenomath}
\begin{equation}
\gamma _R  = \frac{N}{{\sqrt {N^2  - 1} }}\sqrt {1 - \frac{{\varpi ^2 }}{{N^2 }} - 2\frac{\varpi }{N}\frac{{P_z }}{{m_e c}}}
\label{eqA05}
\end{equation}
\end{linenomath}
Taking into account that $\varepsilon_0=-\omega I_R/m_ec^2+\gamma_R=const$, we exclude $P_z$ from the equation for energy $\gamma_R$:
\begin{linenomath}
\begin{equation}
\gamma _R  = \left| {\varpi  \mp \frac{N}{{\sqrt {N^2  - 1} }}\sqrt {1- 2\varepsilon _0 \varpi + \varpi ^2 } } \right|
\label{eqA06}
\end{equation}
\end{linenomath}
Then, we expand Hamiltonian (\ref{eqA01}) around the resonant momentum:
\begin{linenomath}
\begin{eqnarray}
 F &=& \Lambda  + \frac{1}{2}g\left( {I - I_R } \right)^2  + U_w (z,I_R )\sin \zeta , \label{eqA07}\\
 \Lambda  &=&  - \omega I_R  + m_e c^2 \gamma _R ,\quad g = \left. {\frac{{\partial ^2 \gamma }}{{\partial I^2 }}} \right|_{I_R }  = \frac{{\omega ^2 \left( {N^2  - 1} \right)}}{{\gamma _R m_e^2 c^4 }} \nonumber
 \end{eqnarray}
 \end{linenomath}
Hamiltonian $\Lambda(z,P_z)$ describes resonant particle motion in the $(z,P_z)$ plane. We use generating function $S_2=(I-I_R)\zeta+pz$ to introduce new momenta $P_\zeta=I-I_R$, $p=P_z-(\partial I_R/\partial
z)\zeta$, and new coordinate $s=z+(\partial I_R/\partial P_z)\zeta$. Near the resonance $|(\partial I_R/\partial z)\zeta| \ll 1$, $|(\partial I_R/\partial P_z)\zeta| \ll 1$ and we can expand the new Hamiltonian as
\citep[see details in, e.g.,][]{Neishtadt99, Neishtadt11:mmj}
\begin{linenomath}
\begin{eqnarray}
\tilde F &=& \Lambda \left( {P_z ,z} \right) + \frac{1}{2}gP_\zeta ^2  + U_{w,R} \sin \zeta  = \Lambda \left( {p,s} \right) + H_\zeta \nonumber  \\
 H_\zeta   &=& \frac{1}{2}gP_\zeta ^2  - r\zeta  + U_{w,R} \sin \zeta\label{eqA08}\\
  U_{w,R}  &=& \frac{{\sqrt {2\left( {\gamma _R  - \varepsilon _0 } \right)\varpi } }}{{\gamma _R N}}\frac{{eB_w }}{{m_e c\omega }}
 \nonumber
 \end{eqnarray}
\end{linenomath}
Here $r=\{\Lambda, I_R\}_{z,P_z}$ is the Poisson brackets for $\Lambda$, $I_R$. Hamiltonian $H_\zeta$ describes particle motion in the resonant phase plane $(P_\zeta,\zeta)$ with coefficients depending on $(p,s)$ as on parameters. Dynamics of $(p,s)$ is described by Hamiltonian $\Lambda(s,p)$. The coefficient $r$ is \citep[see details in, e.g.,][]{Artemyev18:jpp}
\begin{linenomath}
\begin{eqnarray}
r &=& \left( {\frac{{\partial \Lambda }}{{\partial z}}\frac{{\partial I_R }}{{\partial P_z }} - \frac{{\partial \Lambda }}{{\partial P_z }}\frac{{\partial I_R }}{{\partial z}}} \right)_{\scriptstyle P_z  \approx p \hfill
\atop
  \scriptstyle z \approx s \hfill}  = m_e c^2 \left( {\frac{{\partial \gamma _R }}{{\partial z}}\frac{{\partial I_R }}{{\partial P_z }} - \frac{{\partial \gamma _R }}{{\partial P_z }}\frac{{\partial I_R }}{{\partial z}}}
  \right)  \nonumber\\
  &=& \frac{{m_e c^2 }}{{2\gamma _R }}\frac{{N^2 D}}{{N^2  - 1}}\left( {1 - \gamma _R^2  + \frac{{\gamma _R^2  - \varpi ^2 }}{{N^2 }} + 2\frac{{\left( {\gamma _R  - \varpi } \right)^2 }}{{N^2 }}\frac{{\partial \ln
  N}}{{\partial \ln \varpi }}} \right) \label{eqA09}
 \end{eqnarray}
\end{linenomath}
Here $D=(c/\omega N)\partial \ln\Omega_{ce}/\partial s$ is the ratio of the spatial scales.
To define the probability of trapping, we introduce the area in the $(P_\zeta, \zeta)$ plane filled by trapped particle trajectories, $S_{res}=\oint{P_\zeta d\zeta}$:
\begin{linenomath}
\begin{equation}
S_{res}  = \sqrt {\frac{{8r}}{g}} \int\limits_{\zeta _ -  }^{\zeta _ +  } {\sqrt {a\left( {\sin \zeta _ +   - \sin \zeta } \right) - \left( {\zeta _ +   - \zeta } \right)} \;d\zeta }
\label{eqA10}
\end{equation}
\end{linenomath}
where $a=|U_{w,R}/r|$, and integration limits, $\zeta_{\pm}=\zeta_{\pm}(s)$, are solutions of equations $a\cos\zeta_+=1$ and $a(\sin\zeta_+-\sin\zeta_-)-(\zeta_+-\zeta_-)=0$. For $|D|\ll 1$, the probability of trapping is \citep[see][]{Neishtadt75, Neishtadt99}:
\begin{linenomath}
\begin{equation}
\Pi  = \frac{1}{{2\pi |r|}}\left( {\frac{{\partial \Lambda }}{{\partial s}}\frac{{\partial S_{res} }}{{\partial p}} - \frac{{\partial \Lambda }}{{\partial p}}\frac{{\partial S_{res} }}{{\partial s}}} \right) =
\frac{{\left\{ {\Lambda ,S_{res} } \right\}_{s,p} }}{{2\pi |r|}}
\label{eqA11}
\end{equation}
\end{linenomath}
This probability is small: $\Pi\sim \sqrt{|D|}$. Numerical verifications of Eq.~(\ref{eqA11}) can be found in Fig.~\ref{fig4} and \citet{Artemyev13:pop, Artemyev15:pop:probability}.

\section*{Appendix B: trapped particle acceleration}
In this Appendix we estimate the acceleration rate of trapped particles. We would like to determine the energy change $\Delta\gamma_{trap}$ due to trapping as a function of the initial particle energy and
pitch-angle, $(\gamma, \alpha_{eq})$ (the corresponding pitch-angle change is defined by Eq.~(\ref{eq05})). To be trapped, a particle should resonate with the wave at a location where the probability of
trapping $\Pi>0$. Equation (\ref{eqA11}) shows that $\Pi$ is proportional to the change of $S_{res}$ along the resonant trajectory: $dS_{res}/dt=\partial S_{res}/\partial t + \{\Lambda, S_{res} \}_{s,p}$. Therefore,
particles become trapped at some $S_{res,trap}$ with $dS_{res,trap}/dt>0$.

\begin{figure*}
\includegraphics[width=0.9\textwidth]{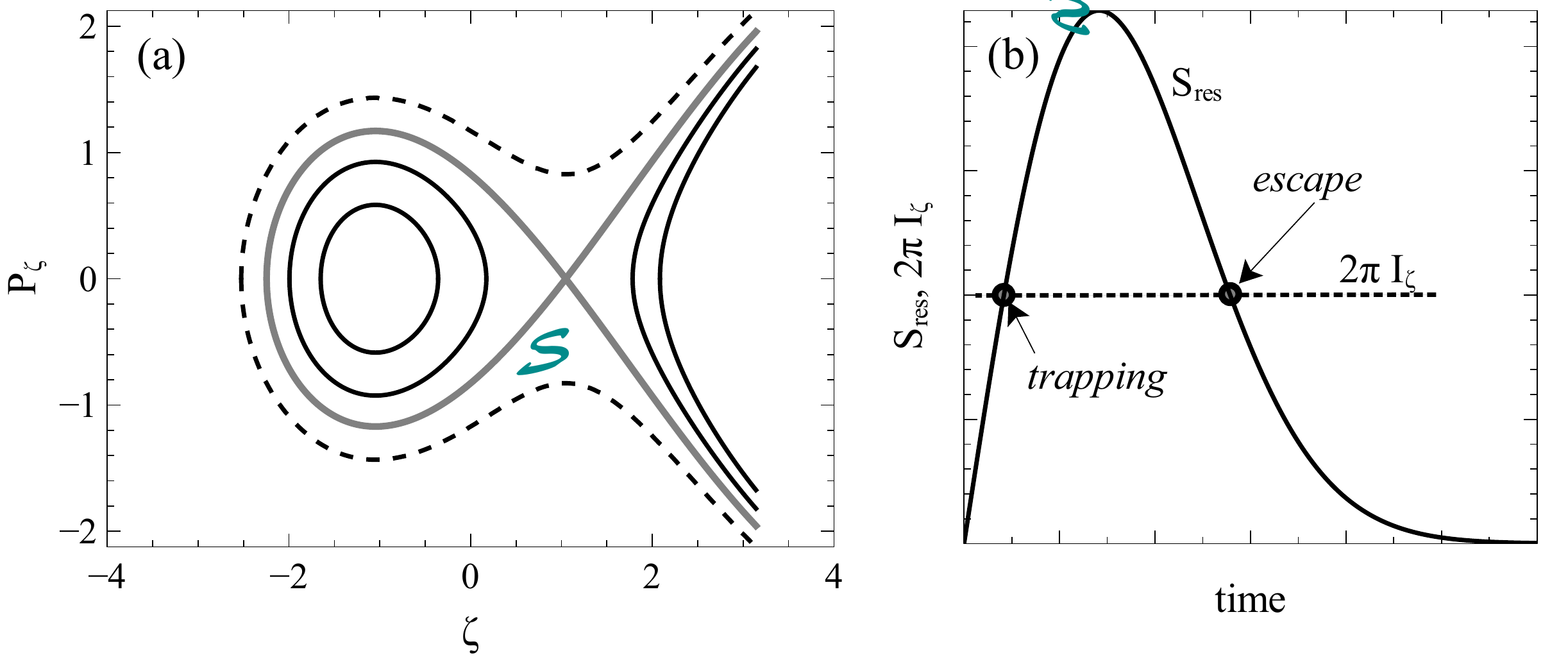}
\centering
\caption{(a) Phase portrait of Hamiltonian $H_\zeta$ given by Eq.~(\ref{eqA08}). Grey color shows the separatrix, closed solid curves are trajectories of trapped particles, the dotted curve is the trajectory of a transient (scattered) particle. (b) A schematic view of the determination of trapping and escaping times. (c) Results of numerical integration of Hamiltonian equations and analytical expressions (\ref{eqA06}), (\ref{eqA10}); see details in the text. (d) A fragment of trapped particle trajectory in the $(\zeta, P_\zeta)$ plane; dimensionless parameter $\chi=\Omega_{ce}(0)R_EL/c$. System parameters are the same as in Fig.~\ref{fig4}.}
\label{figB1}
\end{figure*}

Dynamics of trapped particles are defined by the Hamiltonian $H_\zeta$ in Eq.~(\ref{eqA08}). This is a classical pendulum Hamiltonian with coefficients depending on slowly changing parameters $(s,p)$ determined by
Hamiltonian $\Lambda$ in Eq.~(\ref{eqA08}) (note that the time scale of $(s,p)$ variation is much larger than the time scale of $(\zeta, P_\zeta)$ variations \citep[see details in, e.g.,][]{Neishtadt99,
Artemyev15:pop:probability}). The phase portrait of Hamiltonian $H_\zeta$ is shown in Fig.~\ref{figB1}(a). There are closed trajectories oscillating around the resonance $P_\zeta=0$ (trapped trajectories) confined by the
separatrix $\s$. The area surrounded by the separatrix, $S_{res}$, changes with time (with $(s,p)$) and when this area becomes larger than the area surrounded by the particle trajectory, $2\pi I_\zeta$, particles becomes
trapped  (this determines the condition of $dS_{res}/dt>0$). The area $2\pi I_\zeta=\oint P_\zeta d\zeta$ (integrated along a particle trajectory) is the adiabatic invariant of the system described by Hamiltonian
$H_\zeta$ \citep[e.g.,][]{bookLL:mech} and thus, $I_\zeta$ is conserved during trapped motion. The equation $I_\zeta=const$ defines the condition for particle escape from resonance: $2\pi I_\zeta=S_{res}$ and $dS_{res}/dt <0$. Thus, the position of particle escape from the resonance, $z_{esc}$, is determined by equation $S_{res,trap}=S_{res,esc}$ and $dS_{res,esc}/dt<0$ (see scheme in Fig.~\ref{figB1}(b)). The energy gain of trapped particles is $\Delta\gamma_{trap}=\gamma_R(z_{esc})-\gamma_R(z_{trap})$ with $\gamma_R$ given by Eq.~(\ref{eqA06}). To construct
$\Delta\gamma_{trap}(\gamma,\alpha_{eq})$, we determine the $S_{res}$ profile along the resonant trajectory (Eq.~(\ref{eqA10})) for each $(\gamma,\alpha_{eq})$, define positions of particle trapping and escape, and
calculate the difference of the resonant energy $\gamma_R$ between these positions.

Figure~\ref{figB1}(c,d) compares analytical predictions of trapped particle motion with results of the numerical integration of Hamiltonian equations. Panel (c) demonstrates that trapped particle energy (black curve)
coincides with Eq.~(\ref{eqA06}) (red curve). Particle becomes trapped and escape from the resonance with the same area $S_{res}$ (grey curve). During the trapped motion, the particle oscillates in the $(\zeta, P_\zeta)$
plane (panel (d)) and the area surrounded by the particle trajectory is conserved (blue crosses in panel (c)). This area, $2\pi I_\zeta$, is equal to $S_{res,trap}$ and thus equation $2\pi I_\zeta=S_{res}$ defines the position of particle escape from resonance.

\section*{Appendix C: scattering amplitude}
In this Appendix we estimate the particle energy change due to nonlinear scattering, $\Delta\gamma_{scat}$. The conservation of $\varepsilon_0=\gamma-\omega I/m_ec^2$ shows that $\Delta\gamma=\omega\Delta I/m_ec^2$. To evaluate $\Delta I$, we consider Hamiltonian (\ref{eqA01}):
\begin{linenomath}
\begin{eqnarray}
 \Delta I =  - 2\int\limits_{ - \infty }^{t_R } {\frac{{\partial F}}{{\partial \zeta }}dt}  =  - 2\int\limits_{ - \infty }^{\zeta _ +  } {U_{w,R} \cos \zeta \frac{{d\zeta }}{{gP_\zeta  }}} \label{eqC01} \\
  =  - \sqrt {\frac{{2U_{w,R} }}{g}} \int\limits_{ - \infty }^{\zeta _ +  } {\frac{{\sqrt a \cos \zeta }}{{\sqrt {a\left( {\sin \zeta _ +   - \sin \zeta } \right) - \left( {\zeta _ +   - \zeta } \right)} }}d\zeta }
  \nonumber
 \end{eqnarray}
\end{linenomath}
where $t_R$ is the time of the resonant interaction and we use $dt=d\zeta/\dot\zeta=d\zeta/gP_\zeta$ (see Hamiltonian $H_\zeta$ in Eq.~(\ref{eqA08})). The change $\Delta I$ depends on the phase $\zeta_+$, or alternatively
on $\xi$, where $2\pi\xi = a\sin\zeta_+-\zeta_+$. It can be shown that $\Delta I$ is a periodic function of $\xi$ \citep[see, e.g.,][]{Karpman75PS, Neishtadt99, Artemyev14:pop} and $\xi$ is distributed uniformly for any
reasonable initial set of wave phases $\zeta$ \citep[e.g.,][]{Itin00}. Therefore, to derive the average value of the energy change $\Delta\gamma_{scat}=\omega\langle\Delta I\rangle_{\xi}/m_ec^2$ we need to average $\Delta
I$ over $\xi\in[0, 1]$. An important property of the function $\Delta I(\xi)$ is $\langle\Delta I\rangle_{\xi}=-S_{res}/2\pi$ \citep[see][]{Karpman75PS, Neishtadt99, Dolgopyat12}. Thus, we can express
$\Delta\gamma_{scat}$ as:
\begin{linenomath}
\begin{equation}
\Delta \gamma _{scat}  = \frac{\omega }{{m_e c^2 }}\left\langle {\Delta I} \right\rangle _\xi   =  - \frac{\omega }{{m_e c^2 }}\frac{{S_{res} }}{{2\pi }}
\label{eqC02}
\end{equation}
\end{linenomath}
Equation (\ref{eqC02}) shows that $\Delta\gamma_{scat}$ is about $\sqrt{|D|}\ll 1$, i.e.,  $\Delta\gamma_{scat}$ is of the same order as the probability of trapping $\Pi\sim \sqrt{|D|}$ \citep[see a discussion in][]{Shklyar09:review, Shklyar11:angeo}.


\begin{acknowledgments}
This material is based in part upon work supported by the National Science Foundation under Award No. CMMI-1740777 (D.L.V.). X.J.Z., A.V.A., and V.A. acknowledge NASA contract NAS5-02099 for use of data from the THEMIS Mission, specifically J. W. Bonnell and F. S. Mozer for use of EFI data, A. Roux and O. LeContel for use of SCM data, and K. H.
Glassmeier, U. Auster and W. Baumjohann for the use of FGM data provided under the lead of the Technical University of Braunschweig and with financial support through the German Ministry for Economy and Technology and the
German Center for Aviation and Space (DLR) under contract 50 OC 0302. The work of X.J.Z. was also supported by RBSP-EMFISIS and RBSP-ECT funding 443956-TH-81074 and 443956-TH-79425 under NASA's prime contract No.
NNN06AA01C. We acknowledge the Van Allen Probes EMFISIS data obtained from \url{https://emfisis.physics.uiowa.edu/data/index}, and THEMIS data from \url{http://themis.ssl.berkeley.edu/}. We also thank the World Data
Center for Geomagnetism, Kyoto for providing AE index, and the Space Physics Data Facility at the NASA Goddard Space Flight Center for providing the OMNI data used in this study.
\end{acknowledgments}

\bibliographystyle{agu}

\end{article}

\end{document}